\documentclass[12pt]{article}
\pdfoutput=1
\usepackage{jcapmod}
\usepackage{booktabs}
\usepackage{mathtools}
\usepackage[english]{babel}
\usepackage{amsmath,amssymb,amsbsy,amstext, amsthm}
\usepackage[usenames,dvipsnames]{xcolor}
\usepackage{hyperref}
\hypersetup{
    urlcolor=NavyBlue,
    citecolor=NavyBlue,
    linkcolor=NavyBlue,
}%
\usepackage{array}
\usepackage{url} 
\usepackage{slashed}
\usepackage{multicol}
\usepackage{blkarray}
\usepackage{enumerate}
\usepackage{graphicx}
\usepackage{amsfonts}
\usepackage{amssymb}
  \usepackage{empheq}
  \usepackage{tcolorbox}
\usepackage{xcolor}
\newcommand\myshade{85}
\colorlet{mylinkcolor}{RoyalPurple}
\colorlet{mycitecolor}{WildStrawberry}
\colorlet{myurlcolor}{BlueViolet}
\hypersetup{
  linkcolor  = mylinkcolor!\myshade!black,
  citecolor  = mycitecolor!\myshade!black,
  urlcolor   = myurlcolor!\myshade!black,
  colorlinks = true,
}
\usepackage{colortbl}

\usepackage{float}
\usepackage[utf8]{inputenc}
\RequirePackage{color}
\usepackage[normalem]{ulem}
\usepackage[capitalise]{cleveref}

\newcommand\mathcomma{\,,}
\newcommand\mathperiod{\,.}

\def\dd{\mathrm{d}}

\usepackage{colortbl}
\definecolor{green2}{cmyk}{0, 1, 0.5, 0}
\definecolor{lightgreen}{cmyk}{0.2, 0, 0.2, 0.2}
\definecolor{lightgray}{cmyk}{0.1,0.2,0,0.1}
\definecolor{lightgray2}{cmyk}{0.4,0.4,0,0.8}
\definecolor{black}{cmyk}{1.0,1.0,1.0,1.0}

\allowdisplaybreaks[1]

\usepackage{colortbl}
\definecolor{lightgreen}{cmyk}{0.2, 0, 0.2, 0.2}
\definecolor{lightgray}{cmyk}{0.1,0.2,0,0.1}
\definecolor{lightgray2}{cmyk}{0.1,0.1,0,0.1}

\makeatletter
\newlength{\apb@width}
\newcommand{\autoparbox}[2][c]{\settowidth{\apb@width}{#2}\parbox[#1]{\apb@width}{#2}}

\makeatother

\numberwithin{equation}{section}

\def\ni{\noindent}
\def\be{\begin{equation}}
\def\ee{\end{equation}}

\def\bea{\begin{eqnarray}}
\def\eea{\end{eqnarray}}

\def\lp{\left(}
\def\rp{\right)}

\def\dd{{\rm d}}

\def\Mp{M_{\rm Pl}}

\def\0{{\boldsymbol 0}}

\usepackage{setspace} 

\usepackage{array}
\newcommand{\PreserveBackslash}[1]{\let\temp=\\#1\let\\=\temp}
\newcolumntype{C}[1]{>{\PreserveBackslash\centering}p{#1}}
\newcolumntype{R}[1]{>{\PreserveBackslash\raggedleft}p{#1}}
\newcolumntype{L}[1]{>{\PreserveBackslash\raggedright}p{#1}}

\usepackage{multirow}

\begin{document}

\begin{titlepage}

\setcounter{page}{1} \baselineskip=15.5pt \thispagestyle{empty}

\bigskip\

\vspace{1cm}
\begin{center}

{\fontsize{20}{28}\selectfont  \sffamily \bfseries {Deciphering Coupled Scalar Dark Sectors
 \\}}

\end{center}

\vspace{0.2cm}

\begin{center}
{\fontsize{13}{30}\selectfont Saba Rahimy$^{a,}$\textsuperscript{*}, 
Elsa M. Teixeira$^{b,}$\textsuperscript{†},  Ivonne Zavala$^{a,}$\textsuperscript{‡}
} 
\end{center}

\begingroup
\renewcommand{\thefootnote}{\fnsymbol{footnote}}
\footnotetext[1]{\texttt{saba.rahimy@gmail.com}}
\footnotetext[2]{\texttt{elsa.teixeira@umontpellier.fr}}
\footnotetext[3]{\texttt{e.i.zavalacarrasco@swansea.ac.uk}}
\endgroup

\begin{center}

\vskip 8pt
\textsl{$^a$ Physics Department, Swansea University, SA2 8PP, UK}\\
\textsl{$^b$ Laboratoire Univers \& Particules de Montpellier, \\ CNRS \& Université de Montpellier (UMR-5299), 34095 Montpellier, France}
\vskip 6pt

\end{center}

\vspace{1.2cm}
\hrule \vspace{0.3cm}
\noindent
 {\bf Abstract} \\
 
\ni  Coupled dark sector models have gained significant attention, motivated by recent advances in cosmology and the pressing need to address unresolved puzzles. In this work, we study coupled scalar dark sector models inspired by ultraviolet-complete frameworks such as supergravity and string theory. These models involve scalar couplings arising either from their kinetic terms, through a non-trivial field space metric, or from their scalar potential. We demonstrate how these couplings can be elegantly formulated in terms of an {\em interacting vector}, a standard tool in coupled dark sector studies, and analyse their distinct cosmological effects using a dynamical systems approach. Using this framework, we further investigate an axio-dilaton system recently explored in the literature, where the dilaton also couples to baryons. Intriguingly, we show that certain kinetic and potential interactions may mimic one another or even cancel out, making them observationally indistinguishable. 
If such a distinction becomes possible through observational constraints, it could provide valuable insights into the underlying field space metric and its connection to fundamental physics. 

\vskip 10pt
\hrule

\vspace{0.4cm}
 \end{titlepage}

 \tableofcontents

\section{Introduction}

Interacting dark sector models (IDS), more commonly referred to as interacting dark energy, are a natural extension of the standard $\Lambda$CDM cosmological model in which a direct non-gravitational interaction between dark energy (DE) and dark matter (DM) is allowed. IDS models have a long history\footnote{A list of recent comprehensive reviews is given by e.g.~\cite{Zimdahl:2012mj,Bolotin:2013jpa,Wang:2016lxa,Bahamonde:2017ize,Wang:2024vmw}.}, and one of the main motivations for them is their potential to alleviate the so-called coincidence problem \cite{Comelli:2003cv,Velten:2014nra} -- namely, why the densities of dark energy and dark matter are nearly equal today when they scale so differently during the expansion history of the universe. 
Additionally,  IDS models may reduce fine-tuning of initial conditions by leading both components toward a common scaling solution and have also been explored as a possible way to ease cosmological tensions, such as the Hubble tension\footnote{Note that while coupled dark sector models offer dynamical advantages, they do not address the fundamental cosmological constant problem \cite{Burgess:2013ara}.}  \cite{Yang:2018uae,Guo:2018ans,DiValentino:2019ffd}.

IDS models can generally be classified based on how the interaction is modelled and the properties of the interacting components. For example, one widely studied class involves interacting dark fluids, where dark energy and dark matter are treated as perfect fluids with phenomenologically motivated coupling terms. These class of models are interesting due to their simplicity and flexibility in addressing observational anomalies and cosmic tensions (see e.g.~\cite{Gavela:2009cy,Yang:2017ccc,DiValentino:2017iww,Yang:2018euj,
DiValentino:2019ffd,Yang:2022csz,Yang:2019uzo,Yang:2020uga,Nunes:2022bhn,vandeBruck:2022xbk,DiValentino:2019jae,vanderWesthuizen:2023hcl,Giare:2024ytc,Li:2024qso,Giare:2024smz}).

Alternatively, scalar field models provide a more fundamental description of the dark sector. These include interacting quintessence models, where a scalar field representing dark energy is coupled to dark matter, often modelled as a fluid \cite{Wetterich:1994bg,Amendola:1999er}. A more sophisticated approach involves coupled scalar fields, where both dark energy and dark matter are described by scalar fields that interact either through their potential, their kinetic terms, or both (see e.g.~\cite{Axenides:2004kb, Beyer:2010mt, Marsh:2011gr,Marsh:2012nm,Beyer:2014uja, DAmico:2016jbm,CarrilloGonzalez:2017cll,Carrillo_Gonz_lez_2018,Benisty:2018qed,Brandenberger:2019jfh,Brandenberger:2020gaz,Johnson:2020gzn, Sa:2020fvn, Sa:2021eft, Johnson:2021wou, vandeBruck:2022xbk,Liu:2023kce, Gomes:2023dat,Garcia-Arroyo:2024tqq,Poulot:2024sex,Aboubrahim:2024spa,Smith:2024ayu,Smith:2024ibv} and the reviews \cite{Zimdahl:2012mj,Bolotin:2013jpa,Wang:2016lxa,Bahamonde:2017ize,Wang:2024vmw} for further references). Such scalar models are particularly appealing as they naturally arise in high-energy theories like supergravity and string theory.

In this work, we focus on the latter class of models, examining a nonlinear sigma model with a potential involving two scalars. These scalars represent the dark energy and dark matter components of the universe and interact through a non-trivial field space metric as well as the scalar potential. By studying this system, we aim to elucidate the distinct cosmological effects of potential and kinetic couplings and explore connections to ultraviolet-complete frameworks.
   
The paper is organised as follows: in \cref{sec:SDS}, we introduce the general scalar dark sector model and outline the key cosmological equations governing its evolution. We also demonstrate how both kinetic and potential scalar interactions can be elegantly expressed in terms of an interacting vector, $Q^\nu$,  as commonly used in coupled dark sector models. In \cref{sec:DS}, we employ the powerful method of dynamical systems to analyse the cosmological evolution and properties of the coupled scalar system. Additionally, we apply this approach to an axio-dilaton model where the dilaton is coupled to baryons, offering new insights into the dynamics of such systems. Finally, in \cref{sec:conclu}, we summarise our findings and discuss their broader implications, including potential connections to ultraviolet-complete theories and the prospects for observational constraints.

\section{Interacting scalar fields}\label{sec:SDS}

As anticipated in the introduction, we focus on the case of two scalar fields describing the dark sector, each scalar being fully identified with dark energy and dark matter, respectively. Such a system is described by a non-linear sigma model (NLSM), 
which describes the dynamics of the scalar fields, which take values on a 2D target manifold $\mathcal{M}$ spanned by the scalar fields $\phi^a(x)$. The metric of the target manifold $\mathcal{M}$, is described by $g_{ab}(\phi)$ and its curvature, $R_{\rm fs}$, can be positive, negative or zero. Examples are $\mathcal{M}={\mathbb R}^2 \,\,(R_{\rm fs}=0)$, $\mathcal{M}=S^2 \,\, (R_{\rm fs}>0)$, $\mathcal{M}= {\cal H}^2 \,\, (R_{\rm fs}<0)$, however, the manifold can be more general, and the metric may depend on both scalars.
Moreover, for the two-dimensional case, on which we focus on what follows, the metric can always be put in a {\em conformally} flat form. Although, in general, $g_{ab}$ can depend on the two field coordinates, we consider the case where it depends only on one (combination) of the fields, which is the case in most cases of physical interest, where the system is that of an axion-saxion, as we discuss below.
The most general action we can write for such scalars minimally coupled to gravity is given by \cite{Armendariz-Picon:1999hyi,Langlois:2008mn} (we take for now natural units for which $\Mp=1$)
\be\label{eq:genaction}
S= \int d^4x\sqrt{-g}\left[\frac12 R+ P(X,\phi^a)\right] \mathcomma
\ee
where 
\be
X\equiv -\frac12 g_{ab}(\phi)\partial_\mu\phi^a\partial^\mu\phi^b \mathcomma
\ee
with $a,b=1,\dots n$ for $n$ scalars and $P(X,\phi)$   an arbitrary function of the scalar fields and their kinetic energies. Note that baryons and radiation are not included in the action in \cref{eq:genaction}, as we are interested only in gravitational interactions between all sectors. However, a coupling between the scalar dark sector and baryons can be added, for instance, via a term $S_m(\psi, \tilde g)$ through the conformal metric, $\tilde g$, as e.g.~in the model of \cite{Smith:2024ayu,Smith:2024ibv}, which we consider in \cref{sec:AxioDil}. 

The general action in \cref{eq:genaction}  includes scalar models with both {\em standard kinetic terms} -- characterised by $P(X,\phi) = X- V(\phi)$ --  and more general cases with {\em non-standard kinetic terms}, where the dependence on $X$  is more intricate\footnote{Examples of lagrangians with non-standard kinetic terms include those from k-inflation \cite{Armendariz-Picon:1999hyi,Chiba:1999ka,Armendariz-Picon:2000nqq,Armendariz-Picon:2000ulo} and Dirac-Born-Infeld (DBI) inflation \cite{Silverstein:2003hf,Alishahiha:2004eh,Easson:2007dh}.}.
Multiscalar NLSMs with a potential coupled to gravity naturally emerge in fundamental theories such as supergravity and string theory. In these frameworks, two scalars often originate from a single complex scalar field, $\Phi=\phi+i \chi$. One scalar typically corresponds to an {\em axion}, possessing a continuous shift symmetry that is broken to a discrete one by non-perturbative effects (see \cite{Cicoli:2023opf} for a recent review on string cosmology). 
The second scalar, referred to as the {\em saxion}, often represents a modulus with geometrical interpretation. 
Given this natural pairing, it is compelling to study the dynamics of such scalars together\footnote{Axions and saxions have been considered separately as candidates for dark energy recently in \cite{Bhattacharya:2024kxp}.}. In this work, we focus on the case of standard kinetic terms\footnote{The exploration of non-standard kinetic terms is left for future investigations.}.

\subsection{The scalar's Lagrangian}

We consider the following general action for the axion-saxion
 $(\phi, \chi)$ system
\be\label{eq:standard1}
P(X,\phi) = -\frac{1}{2}\partial_\mu \phi\partial^\mu \phi - \frac{f^2(\phi)}{2}\partial_\mu \chi\partial^\mu \chi  -  V(\phi,\chi)\mathcomma
\ee
that is, the field space metric depends only on the saxion $\phi$, and its curvature is given in terms of the function $f$ and its derivatives  as
\be\label{eq:fscurvature}
R_{\rm fs} = -\frac{2f_{\phi\phi}}{f}\mathcomma
\ee
where $f_{\phi\phi}$ denotes the second derivative of $f$ with respect to $\phi$. Note that for a flat metric with $R_{\rm fs}=0$,   $f_\phi=$const., but $f(\phi)\ne $ const. 
Notice also that the field space metric in \cref{eq:standard1} can be written in various equivalent forms by simple changes of variables: 
\be
\dd s^2=\dd \phi^2 + f^2(\phi) \dd \chi^2 = f^2(r)\left(\dd r^2+ \dd \theta^2\right) = f^2(\rho)\left(\dd \rho^2 +\rho^2 \dd\theta^2 \right)\,.
\ee
As anticipated,  the coupling between the scalars arises due to the kinetic terms -- kinetic interaction --  \textit{via} the metric through the function $f(\phi)$, as well as from the scalar potential $V(\phi,\chi)$ -- potential interaction. 
The case of no kinetic interaction corresponds to $f(\phi)=$const., in which case the only interaction arises through the scalar potential $V$ and is the most popular case in the literature (see e.g.~\cite{Axenides:2004kb,vandeBruck:2022xbk,Liu:2023kce} for some recent models). 
For $f=$const. and a separable potential of the form $V(\phi,\chi)=V_1(\phi)+V_2(\chi)$, there is no interaction between the scalars, so we discard this case. 
Instead, we are interested in the cases where either kinetic,  potential or both interactions are non-trivial. In particular, we aim at disentangling the roles of these interactions\footnote{Specific scalar dark sector models with kinetic and potential interactions have been studied in \cite{Sa:2020fvn, Sa:2021eft,Poulot:2024sex}, while a purely kinetic interaction was considered in \cite{Smith:2024ayu, Smith:2024ibv}. In the following sections, we place these scenarios within a broader framework and analyse them systematically using dynamical systems.}. 

The choice of roles for the scalars ($\phi$, $\chi$) as dark matter or dark energy is driven by phenomenological considerations. For example, since the target space metric is independent of $\chi$, this field has a continuous shift symmetry in the kinetic term. Such symmetry may be preserved, broken mildly or fully in the potential interaction. For example, identifying $\chi$ with an {\em axion} (or more generally, a pseudo-Nambu-Goldstone boson (pNGB)), its shift symmetry may be broken by non-perturbative effects (e.g.~instantons) to a discrete symmetry in the potential. Such symmetry may be further broken spontaneously, generating a monomial-like potential analogous to the axion monodromy mechanism, which was first explored in the context of inflation in \cite{Silverstein:2008sg}.
A natural choice for the roles of the scalars is to identify the axion-like field $\chi$ with dark matter and the saxion-like field $\phi$ with dark energy. We employ this choice in what follows, but the reverse option is also possible.

\subsection{Cosmological equations}

 Let us then consider the action in \cref{eq:genaction} with $P$ given by \cref{eq:standard1}. Focusing on the cosmological evolution, we consider a flat FLRW metric:
 \be
 \dd s^2=- \dd t^2 + a^2(t) \dd x^i \dd x_i\mathcomma
 \ee
 with $a(t)$ the scale factor.  The equations of motion derived from the variation of the action with respect to the metric and the scalars become:
\begin{subequations}\label{eq:EOMs} 
 \begin{align}
     &H^2 =\frac{1}{3}\sum_i\rho_i \mathcomma
     \label{eq:Friedmann}\\
&    \frac{\dot{H}}{H^2} = - \frac{1}{2H^2}\sum_i(\rho_i+p_i) \mathcomma
\label{eq:epsilonfi}\\
     &\ddot\phi^a +3H\dot\phi^a +\Gamma^a_{\,\,bc}\dot\phi^b\dot\phi^c +g^{ab}V_b= 0\mathcomma\label{eq:ScalarsEq}
 \end{align}
\end{subequations}
where  $\rho_i$ is the energy density associated with the species $i$, with $i=\varphi, r, b$ for the scalar field, radiation and baryon contributions, respectively, and we define the scalars' energy density and pressure as
\[\rho_\varphi=\frac{\dot\varphi^2}{2}+V(\phi,\chi) \,, \qquad 
p_\varphi=\frac{\dot\varphi^2}{2}-V(\phi,\chi)\,,\]
with
\be
\dot\varphi^2=g_{ab} \dot \phi^a\dot\phi^b\mathcomma
\ee
 while $\Gamma^a_{\,\,bc}$ are the Christoffel symbols computed from the 2D field space metric $g_{ab}$, and $V_a$ denotes derivative of $V$ with respect to the field $\phi^a$. 
 For the Lagrangian in \cref{eq:standard1}, the scalars' equations of motion in \cref{eq:ScalarsEq} become:
 \begin{subequations}\label{eq:eomscalars}
     \begin{align}
         \ddot \phi + 3H\dot\phi + V_\phi =& ff_\phi\,\dot \chi^2  \mathcomma\\
         \ddot \chi + 3H\dot\chi + \frac{V_\chi}{f^2} =& -2\frac{f_\phi}{f} \,\dot\phi\dot\chi \mathperiod
     \end{align}
 \end{subequations}
 From these equations, it is evident that in the absence of a kinetic coupling (i.e., when the scalars are canonically normalised and $f_\phi=0$), the terms with derivatives of $f$ vanish. In this case, the interaction between the dark sector scalars can only emerge from their potential $V(\phi,\chi)$.   Moreover, it is equally important to recognise that a {\em purely kinetic coupling/interaction} can arise when $f_\phi\ne 0$, even in the absence of any potential interaction, \textit{i.e.}~ $V(\phi,\chi) = W(\phi)+U(\chi)$. Though often overlooked, this distinction is crucial for understanding the interplay between scalar fields in coupled scalar models.
By comparing the effects of kinetic couplings, potential couplings, and their interplay, we aim to determine whether these interactions can leave distinguishable imprints at the cosmological level. Intriguingly, the effects of kinetic and potential interactions can partially or fully cancel one another, potentially mimicking the absence of any coupling. Exploring these scenarios is essential to determine whether detecting the presence and nature of such couplings is feasible based on observational data. 
In the following subsection, we reformulate these cases in a more standard and intuitive framework, employing an interaction vector $ Q^\nu$. 

\subsection{The scalar interaction vector, $Q^\nu$}\label{sec:Q}

In this section, we show how to write the kinetic and potential scalar interactions in terms of an {\em interaction vector}, $Q^\nu$, as commonly used in coupled dark sector models (see e.g.~\cite{Bahamonde:2017ize}). 
Let us first recall that when we have two interacting fluids, the energy density conservation equations can be written as:
 \be\label{eq:intDT}
 \nabla_\mu T_{(1)}^{\mu\nu} = Q^\nu \mathcomma \qquad 
 \nabla_\mu T_{(2)}^{\mu\nu} = -Q^\nu\mathcomma
 \ee
where $Q^\nu$ denotes the interacting vector that gives the energy-momentum exchanged between the two fluids, and such that the {\em total} energy-momentum tensor is conserved, namely:
\be
\nabla_\mu  T_{\rm tot}^{\mu\nu} = \nabla_\mu\lp   T_{(1)}^{\mu\nu} + T_{(2)}^{\mu\nu} \rp= 0 \mathperiod
\ee
We now derive explicitly the interaction vector in the coupled system of scalars for the kinetic coupling $f(\phi)$ and a potential interaction $V(\phi,\chi)$ of the following form:
\be\label{eq:scalarVgen}
V(\phi,\chi) = W(\phi) + g(\phi)U(\chi) 
\mathperiod
\ee
Let us then define the energy density and pressure of the two scalars as follows:
\begin{subequations}\label{eq:rhopNS}
    \begin{align}
        &\rho_\phi =\frac{\dot\phi^2}{2} + W(\phi)\mathcomma \qquad \quad 
        p_\phi =\frac{\dot\phi^2}{2} - W(\phi)\mathcomma\label{eq:rhopfi}\\
        &\rho_\chi =\frac{f^2\dot\chi^2}{2} + g(\phi)U(\chi) \mathcomma \qquad 
        p_\chi =\frac{f^2\dot\chi^2}{2} - g(\phi)U(\chi) \mathperiod\label{eq:rhopchi}
    \end{align}
\end{subequations}
With these definitions, we can write \cref{eq:eomscalars} as follows:
\begin{subequations}\label{eq:eomscalars2}
    \begin{align}
       \dot\rho_\phi +3H(\rho_\phi + p_\phi) = &\frac{f_\phi}{f} \lp\rho_\chi + p_\chi\rp \dot\phi -\frac{g_\phi}{2g}(\rho_\chi - p_\chi)\,\dot\phi  \mathcomma\\
        \dot\rho_\chi +3H(\rho_\chi + p_\chi) = & -\frac{f_\phi}{f} \lp\rho_\chi + p_\chi\rp \dot\phi  +\frac{g_\phi}{2g}(\rho_\chi - p_\chi)\,\dot\phi \mathperiod
    \end{align}
\end{subequations}
That is, the interaction vector can be written as
\be\label{eq:KPIntQ0}
Q^0 = \frac{f_\phi}{f} \lp\rho_\chi + p_\chi\rp \dot\phi-\frac{g_\phi}{2g}(\rho_\chi - p_\chi)\,\dot\phi   \mathcomma
\ee
for a homogeneous scalar field $\phi$, or more generally:
\be\label{eq:KPIntQ}
Q^\nu =\frac{f_\phi}{f} \lp\rho_\chi + p_\chi\rp \nabla^\nu\phi -\frac{g_\phi}{2g}(\rho_\chi - p_\chi)\,\nabla^\nu\phi\,.
\ee

%
\smallskip

\ni This analysis demonstrates that both the kinetic and potential interactions can be elegantly expressed using the commonly adopted formalism of an interacting vector.  
It is interesting that, as evident from the scalar interaction vector in \cref{eq:KPIntQ}, the interaction arises from the kinetic energies with a coupling dictated by the field space metric, $f(\phi)$, as well as from the potential energies, with a coupling determined by $g(\phi)$. 
We also see clearly how to turn the kinetic and potential couplings on and off, which are controlled independently by $f$ and $g$. 
Note also that at this stage, we have not specified which field plays the role of DM and which DE, so we can choose the potentials for the scalars to be of the suitable form, depending on what roles they play.  

Furthermore, the energy exchange in the dark sector can delay or advance the onset of dark energy dominantion, depending on the nature and sign of the coupling. As will be discussed in \cref{sec:DS}, the presence of scaling attractors can lead to cosmic histories in which acceleration emerges naturally at a time determined by the coupling parameters, rather than being fixed by initial conditions or fine-tuned parameters. This offers a possible dynamical resolution to the ``why now?” conceptual problem of dark energy (why the acceleration of the universe is occurring at very low redshifts) by linking the onset of acceleration to the characteristic dynamics of the dark sector.

\section{Dynamical Systems Framework for Scalar Interactions}\label{sec:DS}

In this section, we analyse the system of coupled scalars using a dynamical systems approach. As explained earlier, for minimally coupled scalars, the interaction can originate from the potential term, the kinetic term, or a combination of both in the Lagrangian. To systematically investigate these possibilities, we adopt an averaging approach akin to the one in \cite{Sa:2021eft}, where the dark matter field $\chi$ is assumed to oscillate around the minimum of its potential, effectively behaving like a non-relativistic fluid. This is done to examine the effects of a pure kinetic coupling, a pure potential coupling, and scenarios where both couplings are simultaneously active.
The need for an averaging approach arises because, when modelling dark matter as a scalar field, its matter-like behaviour emerges due to oscillations around the minimum of its potential. These oscillations are not straightforward to capture using a standard dynamical systems analysis or even direct numerical simulations. By averaging these oscillations, we can extract the system's effective behaviour and uncover the subtle contributions of the couplings. 
 Our aim is to isolate and analyse the impact of these couplings, compare their effects, and assess whether their presence - or absence - could manifest in observable cosmological signatures. This is a key question with far-reaching implications, as the ability to distinguish between kinetic and potential couplings observationally would provide invaluable insights into the underlying physics of the scalar field interactions and their nature and origin. 

\subsection{Dynamical variables}\label{subsec:variables}

We now make a concrete choice for the roles of the scalars in \cref{eq:eomscalars}. Specifically, we identify dark energy with the ``saxion"-like field, $\phi$, and dark matter with the ``axion"-like field, $\chi$. We then follow   \cite{Sa:2021eft} to study the dynamical system, assuming an averaging over dark matter's kinetic and potential energy. 
We start with the general scenario where both the kinetic and potential coupling are active, and besides radiation, we also include baryons. Although baryons are usually not included in a dynamical system analysis due to the present-day value of $\Omega_{b,0} \approx 0.05$, its relevance lies in the fact that baryons made a non-negligible contribution to the total matter budget during the matter-dominated era playing an important role in shaping the whole dynamics of the universe. Since we want to model dark matter only by the axion field $\chi$, it is essential to include this effect explicitly.

Let us  define the following dynamical variables:
\be\label{eq:variables1}
x=\frac{\dot \phi}{\sqrt{6}H} \mathcomma\qquad y=\frac{\sqrt{W}}{\sqrt{3}H} \mathcomma\qquad v=\frac{\sqrt{\rho_r}}{\sqrt{3}H}\mathcomma\qquad u=\frac{\sqrt{\rho_b}}{\sqrt{3}H}\mathcomma
\ee
and 
\be\label{eq:variables2}
\beta=\frac{f_\phi}{f}\mathcomma\qquad  \lambda = -\frac{W_\phi}{W}\mathcomma\qquad \gamma= -\frac{g_\phi}{g}\mathcomma
\ee
for which the Hubble constraint, \cref{eq:Friedmann}, becomes
\be\label{eq:Hconstr}
\Omega_\chi = 1-x^2-y^2 - v^2-u^2\mathcomma 
\ee
where we defined the energy density parameter of DM as
\be
\Omega_\chi = \frac{\rho_\chi}{3H^2}\mathperiod
\ee
Since we take the axion-like field to be dark matter, its average equation of state has to be $w_\chi\approx 0$. 
This behaviour can be achieved for the dark matter field, $\chi$, oscillating rapidly around its minimum\footnote{This behaviour remains valid in the case of a coupled potential $g(\phi) U(\chi)$, as the rapid oscillations of the axion are still dictated by $U(\chi)$, effectively justifying the approximation despite the coupling.} such that it behaves as a non-relativistic dark-matter fluid with 
equation of state $\langle p_\chi \rangle =0$, where $\langle\rangle$ denotes the average over an oscillation period. Then, from \cref{eq:rhopchi}, we have that 
$\langle\dot\chi^2\rangle = \rho_\chi f^{-2}$, $\langle U\rangle = \frac{\rho_\chi g^{-1}}{2}$. Using this, we  get 
\be
\frac{f^2\langle\dot\chi^2\rangle }{3H^2} =\Omega_\chi\mathperiod
\ee
We then consider $f(\phi)\ne $const.~and $g(\phi)\ne $const.~in \cref{eq:KPIntQ} for which the interaction vector in \cref{eq:KPIntQ0} takes the form
\be\label{eq:Qfinal}
Q^0=\left(\frac{f_\phi}{f} - \frac{g_\phi}{g}\right)\,\rho_\chi\,\dot\phi\mathperiod
\ee
\smallskip
\ni This takes a similar form of a coupled quintessence model; however, here, both dark matter and dark energy are described by scalar fields, and both kinetic and potential couplings appear as dictated by the interaction vector in \cref{eq:KPIntQ0}. Indeed,  we can turn on and off the kinetic and potential couplings in \eqref{eq:Qfinal} by setting either  $f(\phi)$ or $g(\phi)$ to a constant\footnote{In \cite{Sa:2021eft}, a potential of the form  $V(\phi,\chi) = W(\phi)(V_0+U(\chi))$ was considered. Thus, in that case, setting $W(\phi)=1$ leaves a pure cosmological constant and no scalar potential for dark energy.}. 

Finally, we  also have the energy density parameters for radiation, baryons and the scalar field given by 
\be
\Omega_r = v^2\mathcomma \qquad \Omega_b = u^2\mathcomma \qquad \Omega_\phi = x^2+ y^2\mathperiod  
\ee
We also  write down expressions for the dark energy equation of state as well as the effective one in terms of these variables as follows:
\be
\frac{p_\phi}{\rho_\phi}= w_\phi =\frac{x^2-y^2}{x^2+y^2} \mathcomma 
\ee
and
\be
\frac{p_T}{\rho_T}=w_{\rm eff} = x^2 -y^2 +\frac{1}{3}v^2\mathcomma
\ee
where $p_T=\sum_i p_i$ and similarly for $\rho_T$. On the other hand, acceleration is determined by the condition $\ddot a>0$, which can be written in terms of the derivative of the Hubble parameter as:
\be
\epsilon\equiv -\frac{\dot H}{H^2} = \frac{3}{2}(1+w_{\rm eff}) \mathcomma
\ee
with acceleration of the expansion corresponding to $w_{\rm eff}<-1/3$.

In terms of the variables in \cref{eq:variables1,eq:variables2}, the equations of motion in \cref{eq:EOMs} become 
\begin{subequations}\label{eq:gen_dynsys_rad_bar}
    \begin{align}
        x'=& -3\,x +3 \,x^3 +\frac{3}{\sqrt{6}} \lambda \,y^2+  2 \,x \,v^2   +\frac32 x\, u^2 +\frac32 x\,\Omega_\chi + \sqrt{\frac{3}{2}}  \, \frac{\lp 2 \,\beta  +\gamma  \rp}{2}\,\Omega_\chi \mathcomma \\
        y'= & y \left( -\frac{\sqrt{6}}{2} \lambda x + 3\, x^2 +2v^2 + \frac32 u^2 +\frac32 \Omega_\chi \right)\mathcomma\\
        v'= & v \left( -2 + 3x^2 + 2v^2 +\frac32 u^2 +  \frac{3}{2} \Omega_\chi \right)\mathcomma\\
        u'= & u \left( -\frac32 + 3 x^2 + \frac{3}{2} u^2 +2 v^2 +  \frac{3}{2} \Omega_\chi  \right)\mathcomma 
        \end{align}
\end{subequations}
where $'=d/dN$ with $N=\log a$. 
To close the system, we also need to express the derivatives field space metric and the potential couplings appearing in the field equations, which give the extra equations:
\be\label{eq:abg}
        \beta'=\sqrt{6} x\lp \frac{f_{\phi\phi}}{f} - \frac{f_\phi^2}{f^2}\rp\mathcomma \quad
        \lambda'= -\sqrt{6} x\lp \frac{W_{\phi\phi}}{W} - \frac{W_\phi^2}{W^2}\rp \mathcomma\quad
     \gamma'= -\sqrt{6} x\lp \frac{g_{\phi\phi}}{g} - \frac{g_\phi^2}{g^2}\rp \mathperiod
\ee
The system is thus, in principle, eight-dimensional. However, from now on, we focus on the case where the couplings are constant such that the interaction vector becomes:
\be\label{eq:Qsimple}
Q^0=\frac{\left(2\beta+\gamma\right)}{2}\,\rho_\chi\,\dot\phi\mathcomma
\ee
implying
\be\label{eq:functions}
f (\phi) = f_0 e^{\beta \phi}\mathcomma \quad W (\phi) = W_0 e^{-\lambda \phi}\mathcomma \quad \text{and}\quad g (\phi) = g_0 e^{-\gamma \phi}\mathcomma
\ee
where $f_0$, $W_0$ and $g_0$ are scaling parameters ($W_0$ has dimensions of mass to the fourth while $f_0$ and $g_0$ are dimensionless) that do not enter the dynamical system.
With this choice, $\beta'=\lambda' =\gamma'=0$ and thus, the relations in \cref{eq:abg} are trivially satisfied. This reduces the system to a four-dimensional one, which is more tractable. 
We notice here that even if the coupling constants $\beta$ and $\gamma$ have different origins and fundamental roles, they enter the dynamical equations (at least at the background level) in the same way and thus become degenerate. For this reason, hereafter, we introduce the following redefinition
\begin{equation}\label{eq:xidef}
    \xi \equiv 2 \beta +\gamma \mathcomma
\end{equation}
which fully captures the coupling dynamics and reduces the number of free parameters to two: $\{ \lambda, \xi \}$. This means that when $\xi=0$, either the two couplings vanish or they cancel each other in the equations. In terms of this parameter, the coupling takes the form often used in coupled quintessence models (see e.g.~\cite{Wang:2016lxa,Bahamonde:2017ize,Wang:2024vmw}). However, we should keep in mind that this simplification arises due to our choice of $f, g$ above. It is evident, on the other hand, that our general interaction vector \eqref{eq:KPIntQ} includes more general situations. 

The physical phase space can be restricted by considering that the relative energy density of each fluid is always positive and normalised. Given that $\Omega_{\phi} =x^2 + y^2$, $\Omega_{r} =v^2$ and $\Omega_{b} =u^2$, this translates into the conditions: $0 \leq x^2 + y^2 \leq 1$, $0 \leq v^2 \leq 1$ and $0 \leq u^2 \leq 1$. The first condition defines a unitary circle in the $(x,y)$-plane, centred at the origin where $\Omega_{\phi}=0$. On the other hand, points on the unit circle correspond to scalar field dominated solutions where $\Omega_{\phi}=1$.

Examining the dynamical system of equations in \cref{eq:gen_dynsys_rad_bar}, we conclude that $y=0$, $v=0$ and $u=0$ are invariant sets of the system, which is furthermore invariant under the transformations $y \rightarrow -y$, $v \rightarrow -v$ and $u \rightarrow -u$, implying that, without loss of information, we can focus solely on positive values of $y$, $v$ and $u$. This means the physical phase space in the $(x,y)$-plane is reduced to the positive-$y$ half-unit disk for the scalar field variables.

Moreover, the system is invariant under the simultaneous transformation $(x,y,\lambda,\xi) \rightarrow (-x,-y,-\lambda,-\xi)$. This implies that the phase space is fully described if we consider only non-negative values of one of the parameters, which we choose to be $\lambda$. Summarising, we have a four-dimensional phase space, defined in the region $$-1 \leq x \leq 1,\ 0 \leq y \leq 1, \  0 \leq v \leq 1, \ 0 \leq u \leq 1 \, ,$$ and two free parameters $$\xi \in \mathbb{R}\ \text{and}\ \lambda \in \mathbb{R}_{\geq 0}\, ,$$ to describe the system fully. We are now interested in the fixed points (or stability or critical points) of the system where the trajectories may stay constant, that is, at a fixed point $(x,y,v,u) = (x_c,y_c,v_c,u_c)$ we have  $(x', y', v', u')=(0,0,0,0)$. 
The advantage of looking for fixed point solutions is that, even without describing the entire expansion history of the universe, we can study the well-defined asymptotic state towards which the system will evolve depending on the parameter space of the model in consideration. We also examine the relevant cosmological parameters at each fixed point ($w_{\rm eff}$ and $\Omega_{\chi}$), listing the parameter regions in which each fixed point exists and corresponds to an accelerated expanding solution.
Finally, we perform a stability analysis of each fixed point by studying the eigenvalues of the Jacobian matrix of the system in \eqref{eq:gen_dynsys_rad_bar}. In general, we present the stability character of each fixed point in terms of the parameters $\{\lambda,\xi\}$. Whenever the expressions are too intricate, we refer to the illustration of the stability regions in a contour plot.

\subsubsection{Fixed points and their properties}

\begin{figure}[H]
    \centering
    \includegraphics[width=0.45\linewidth]{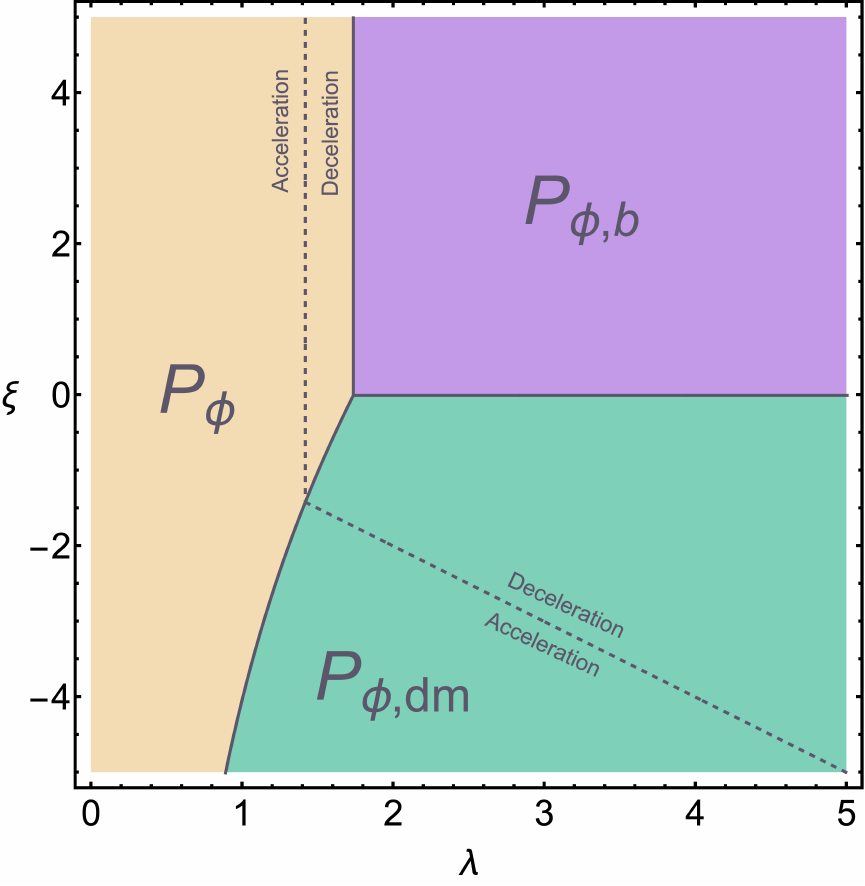}
    \caption{Illustration of the parameter space where each fixed point of the system in \cref{eq:gen_dynsys_rad_bar} as listed in \cref{tab:gen_dynsys_rad_bar} is an attractor according to the study in \cref{tab:Stabgen_r_b}. The dashed lines delimit the region where there is acceleration at the respective fixed point.}\label{fig:reg_stab}
\end{figure}

The fixed points of the system in \cref{eq:gen_dynsys_rad_bar}  are summarised in \cref{tab:gen_dynsys_rad_bar} and their corresponding stability in \cref{tab:Stabgen_r_b}. The properties of these points are as follows: 

\begin{enumerate}[{\bf 1.}]
    \item $P^\pm_{\rm k}$, {\em Kination domination}. These points are dominated by the kinetic energy of the dark energy scalar $\phi$, with $x=\pm1$, and thus $w_{\rm eff}=1$, meaning that the acceleration is never realised in this regime. They are independent of the parameters and always present in the phase space. These are fully unstable points for suitable values of $\xi, \lambda$ as summarised in \cref{tab:Stabgen_r_b}. Being the only repellers of the system and having complementary stability regions, one of these points will be the past attractor for all the possible phase space trajectories. These are standard points in quintessence models where the energy density of the universe is dominated by the scalars' kinetic energy with $\rho_{\rm k}\propto a^{-6}$ and $a(t)\propto t^{1/3}$. 
    
    \item $P_r$, {\em Radiation domination}. This point has $w_{\rm eff}=1/3$, and it always exists in the phase space. It is a saddle for all values of the parameters as indicated in \cref{tab:Stabgen_r_b}. The universe expands dominated by radiation with $\rho_r\propto a^{-4}$ and $a(t)\propto t^{1/2}$.

    \item $P_b$, {\em Baryon domination}. This point has  $w_{\rm eff}=0$, and it is a saddle for all values of the parameters as indicated in \cref{tab:Stabgen_r_b}. The universe expands fully dominated by baryonic matter with $\rho_b\propto a^{-3}$ and $a(t)\propto t^{2/3}$. 

    \item $P_{{\rm k},dm}$, {\em Kination-(dark)matter scaling}. At this fixed point, both the kinetic energy of dark energy and the energy density of the dark matter scalar dominate with $w_{\rm eff} = \frac{\xi^2}{6}$ and $\Omega_\chi=1-\frac{\xi^2}{6}$. The scale factor at this point evolves as $a(t)\propto t^{2/3(1+w_{\rm eff})}$. Thus, it can give rise to (dark) matter domination for $\xi$  sufficiently small. In particular, for $2\beta=-\gamma$ (recall the definition of $\xi$ in  \eqref{eq:xidef}), this point mimics exactly  (dark) matter domination, even if both kinetic and potential couplings are present. This point is a saddle and coincides with $P_{\rm k}^+$ for $\xi=\sqrt{6}$.

    \item $P_{{\rm k},r,dm}$, {\em Kination-radiation-(dark)matter scaling}. This is an interesting point which evolves as if it was dominated by radiation, even though radiation, scalar kinetic energy and dark matter contribute ($(x,v,\Omega_\chi)\ne0$). This has $w_{\rm eff}=1/3$ and it is a saddle. For $\xi=\sqrt{2}$ this point coincides with $P_{{\rm k},dm}$ for $\xi=\sqrt{6}$.
    
    \item $P_{\phi}$, {\em Scalar domination}. Familiar to quintessence models where both kinetic and potential energy of dark energy dominate\footnote{For this point, we consider $(\xi - 2 \lambda)\ne0$.}. This point can have accelerated expansion for $\lambda <\sqrt{2}$, for which it is also an attractor. For $\lambda=\sqrt{6}$ this point coincides with $P_{\rm k}^+$.
    
    \item $P_{\phi,dm}$, {\em (Dark)Matter scaling}. Both kinetic and potential energy of the dark energy scalar are non-zero with $w_{\rm eff}=-\frac{\xi}{\xi-2\lambda}$ and $\Omega_\phi, \Omega_\chi\ne0$. Thus for $2\beta=-\gamma$, $w_{\rm eff}=0$ and the universe evolves as if it was completely dominated by (dark) matter, even if both kinetic and potential couplings may be present. In this limit,  $\Omega_\chi=1-\frac{3}{\lambda^2}$ and this point is equivalent to the standard matter scaling point in uncoupled quintessence models. The non-trivial couplings  introduces a non-vanishing $w_{\rm eff}$ and thus a region of values for $\xi, \lambda$ where an accelerating universe  is possible
(see e.g.~\cite{Bahamonde:2017ize}). We summarise the regions of acceleration and deceleration for this point in \cref{fig:reg_stab}. For some values of the couplings, $\xi,\lambda$, this point can be an attractor or a saddle as summarised in \cref{fig:reg_stab}. For $\lambda=\sqrt{6}$ and $\xi=\sqrt{6}$ this point coincides with $P_{\rm k}^+$.
    
    \item $P_{\phi,r}$, {\em Radiation scaling}. Both the scalar and radiation have a non-zero contribution with $w_{\rm eff}=1/3$. Thus, the universe evolves as if it were dominated entirely by radiation. This point is a saddle. For $\lambda=2$  this point coincides with $P_\phi$.

\item $P_{\phi,b}$, {\em Baryon scaling}. This critical point is equivalent to the matter-scaling where both DE scalar and baryons have a non-zero contribution, $\Omega_\phi, \Omega_b \ne 0$. Still, the universe evolves as it was purely dominated by baryonic matter with an effective equation of state parameter $w_{\text{eff}} = 0$. For $\lambda=\sqrt{3}$  this point coincides with $P_\phi$.
\end{enumerate}

\begin{table}[H]
    \centering
\resizebox{17cm}{!}{
\begin{tabular}{|c|c|c|c|c|c|}
\hline
\rowcolor{gray!30}   $(x,y,v,u)$ &  $w_{\rm eff}$ & Existence & $\Omega_\chi$ & Accelerating \\
    \hline 
     $P_{\rm k}^{\pm}=(\pm 1, 0,0,0)$ &  1  & $\forall \,\xi, \lambda $ & 0 & no\\
     \hline 
     $P_r=(0, 0, 1,0)$ &   1/3  & $\forall \,\xi, \lambda $ & 0 & no\\
     \hline
   $P_{b} =\lp 0,0,0,1 \rp$ &   $0$  & $\forall \,\xi, \lambda $ & $0$ & no \\
     \hline 
      $P_{{\rm k},dm}=\lp \frac{\xi}{\sqrt{6}}, 0,0,0\rp$ &  $\frac{\xi^2}{6}$  & $\xi^2 \leq 6$, \,\,$\forall\lambda$ & $1-\frac{\xi ^2}{6}$ & no \\
     \hline
     $P_{{\rm k},r,dm}=\lp \sqrt{\frac23}\frac{1}{\xi}, 0,\sqrt{1-\frac{2}{\xi ^2}},0\rp$ &  $1/3$  & $\xi^2 \geq 2$, \,\,$\forall\lambda$ & $\frac{4}{3 \xi ^2}$ & no \\
     \hline
      $P_\phi=\lp \frac{\lambda}{\sqrt{6}}, \sqrt{1-\frac{\lambda^2}{6}},0,0 \rp$&     $\frac{\lambda^2}{3}-1$  & $0 \leq\lambda\leq \sqrt{6}$, \,\,$\forall\xi$ & 0 & 
      for $0\leq\lambda<\sqrt{2}$ \\
      \hline 
  $P_{\phi, dm} =\lp \frac{\sqrt{6}}{2 \lambda -\xi}, \frac{\sqrt{\xi ^2-2 \lambda  \xi +6}}{\sqrt{(\xi -2 \lambda )^2}}, 0 ,0  \rp$ &   $\frac{\xi }{2 \lambda -\xi }$  & $0<\lambda \leq \sqrt{6}\,\land\, \xi \lambda \leq 2\lambda^2-6$ \, $\lor$  
     &  $\frac{4 \lambda ^2-2 \lambda  \xi -12}{(\xi -2 \lambda )^2}$   & $0<\lambda \leq \sqrt{2}\land \xi \leq \frac{2 \lambda ^2-6}{\lambda}$  $\lor$ \\ &&
     $\lambda >\sqrt{6}\, \land\, \xi \leq \lambda -\sqrt{\lambda ^2-6}$  &&$\lambda >\sqrt{2}\land \xi <-\lambda$
     \\
  \hline
   $P_{\phi, r} =\lp 2\sqrt{\frac{2}{3}} \frac{1}{\lambda}, \frac{2}{\sqrt{3}\lambda}, \sqrt{1-\frac{4}{\lambda^2}} ,0 \rp$ &  $1/3$  & $\lambda\geq 2$,  $\forall\, \xi$ & $0$ & no \\
  \hline
   $P_{\phi, b} =\lp \sqrt{\frac{3}{2}} \frac{1}{\lambda}, \sqrt{\frac{3}{2}}\frac{1}{\lambda}, 0, \sqrt{1-\frac{3}{\lambda^2}}    \rp$ &   $0$  &  $\lambda\geq \sqrt{3}, \forall\, \xi$  & $0$ & no  \\
  \hline
\end{tabular}
}
    \caption{Fixed points for the system in \cref{eq:gen_dynsys_rad_bar} (recall definition of $\xi$ in \eqref{eq:xidef}  and that $\lambda\geq0$). }
    \label{tab:gen_dynsys_rad_bar}
\end{table}

\begin{table}[H]
    \centering
\resizebox{\textwidth}{!}{
\resizebox{16cm}{!}{
\begin{tabular}{|c|c|}
\hline
\rowcolor{gray!30}   Fixed Point & Stability \\
    \hline 
     $P_{\rm k}^{\pm}$ &  $P_{\rm k}^{-}$: Repeller for $\xi >-\sqrt{6}\land \lambda \geq 0$; Saddle otherwise\\
     &  $P_{\rm k}^{+}$: Repeller for $\xi <\sqrt{6}\land 0\leq \lambda <\sqrt{6}$; Saddle  otherwise \\
     \hline 
     $P_r$ & Saddle   \\
     \hline 
     $P_b$ & Saddle   \\
     \hline 
      $P_{{\rm k},dm}$ &   Saddle  \\
     \hline
     $P_{{\rm k},r,dm}$  &  Saddle   \\
     \hline
      $P_\phi$ & Attractor for $\lambda =0\lor \left(0<\lambda <\sqrt{3}\land \xi >\frac{2 \lambda ^2-6}{\lambda }\right)$; Saddle otherwise \\
      \hline
  $P_{\phi, dm}$ & Attractor for $0<\xi <\frac{2 \lambda ^2-6}{\lambda }$; Saddle otherwise\\
  \hline
   $P_{\phi, r} $ &  Saddle  \\
      \hline
  $P_{\phi, b}$ & Attractor for $\lambda > \sqrt{3} \land \xi >0$; saddle otherwise \\
  \hline
\end{tabular}
}}
    \caption{Stability of the fixed points for the system in \cref{eq:gen_dynsys_rad_bar}.}
    \label{tab:Stabgen_r_b}
\end{table}

The stability parameter regions for the regular fixed points are completely disjoint and complementary. This implies that for a specific set of parameters ${\lambda,\xi}$, there exists a unique attracting fixed point in the system, as illustrated in \cref{fig:reg_stab}, where different colours represent the variously labelled attractors.

\subsection{Realistic cosmologies }

In this section, we investigate cosmological solutions to the coupled scalar dark sector discussed above, which can reproduce the known evolution of the universe at the background level. 
Independently of the coupling origin (potential or kinetic), a realistic cosmology should have a past epoch of radiation followed by matter domination lead by a combination of baryons and scalar dark matter, $\chi$, to finally reach accelerated expansion driven by the scalar quintessence field $\phi$.
From the description of the fixed points above, we see that a realistic solution should start from the radiation fixed point $P_r$ to then follow to $P_{{\rm k},dm}$, passing close to the baryonic point $P_{b}$, to finally reach an accelerated epoch dictated by the condition $w_{\rm eff}=x^2-y^2+\frac13 v^2<-1/3$. The particular trajectory will depend on the initial conditions and phenomenological requirements. We know that the kination critical points can be past repellers or saddles, with $\rho_{\rm kin}\propto a^{-6}$. Thus, far in the past, these will eventually dominate. On the other hand, we know that the epoch of radiation domination should start by the onset of Big-Bang Nucleosynthesis (BBN), that is, at least around 20 e-folds before today with\footnote{Here $1+z=a_0/a$ denotes cosmological redshift.} $z_{\rm BBN}\approx 4\times 10^8$. On the other hand, radiation-matter equality happened  around $z_{r=m}\approx 3400$, or $N_{r=m}\approx -8.1$. Finally, the present epoch of accelerated expansion is given by $z_{m={\rm DE}}\approx 0.3$, or $N_{m={\rm DE}}\approx -0.26$. So, any realistic solution should account for the correct duration of these epochs. Moreover, a realistic solution should be one in which the values of the energy density parameters and effective equation of state match today's values.
Although, in general, observational constraints are model-dependent, we do not expect large changes in their values. Therefore, we take as fiducial values for concrete, realistic solutions the following:
\be\label{eq:EDs0}
\Omega_{b,0}\approx 0.05 \mathcomma\quad \Omega_{\chi, 0}\approx 0.26 \mathcomma\quad \Omega_{\phi,0}\approx 0.685 \mathcomma\quad \Omega_{r,0} \approx 10^{-4}\mathperiod
\ee
We can also determine the maximum value that $\Omega_\chi$ and $\Omega_b$ should reach during the matter-dominated phase. This can be estimated using the fiducial values of today's energy density parameters above\footnote{This can be done by considering the expressions for $\Omega_n(a)$, assuming $\rho_\phi / \rho_{\phi,0} \sim 1$  and neglecting derivatives of this quantity with respect to $ a $ during matter domination \cite{Andriot:2024jsh}.}  \cite{Andriot:2024jsh}, which gives $\Omega_b^{\text{max}} \approx 0.16$ and  $\Omega_\chi^{\text{max}} \approx  0.84$. 
From this constraint, we can, in principle, derive an upper bound on $\xi$, since $P_{{\rm k}, dm}$ has $\Omega_{\chi} = 1 - \xi^2/6$, leading to $\xi^2 \lesssim 0.96$, which in turn gives $w_{\rm eff}^{\text{max}} \lesssim 0.16$. However, this estimate neglects the scalar contribution, which results in an overestimate—allowing too large a value of $\xi$, leading to a non-negligible $\Omega_{\phi}$ in the past. This violates the assumptions behind the estimate and thus necessitates smaller values of $\xi$, as we will see.
On the other hand, for $(2\beta = -\gamma)$, the contributions from the kinetic and potential couplings cancel, mimicking a standard evolution even in the presence of nonzero couplings, as anticipated earlier.

Another important constraint comes from today's acceleration. Although it is possible to have realistic cosmologies featuring transient acceleration for an exponential potential as the one we are considering at the moment for dark energy, this possibility seems to be excluded with a bound on $\lambda\lesssim  0.537$ \cite{Akrami:2018ylq,Bhattacharya:2024hep}\footnote{The bounds depend a little on the data used and we refer the reader to \cite{Bhattacharya:2024hep} for a detailed analysis.}. We do not expect a significant change of this result in the coupled case; however, in the coupled case, transient acceleration for values of $\lambda\gtrsim \sqrt{2}$ which are expected by recently discussed quantum gravity constraints (see e.g.~\cite{Rudelius:2021oaz}) may be possible with a universe which evolves towards the matter scaling point $P_{\phi,dm}$. We show an example of this below. 

On the other hand, coupled dark sector models are attractive due to their potential to address the coincidence problem (why is the energy density of matter and dark energy of around the same order today). In the standard case of coupled quintessence, this arises due to the scaling accelerating solution, also present in our case, $P_{\phi,dm}$. However, when the scaling solution allows for acceleration, the matter fixed point $P_{k,dm}$ does not exist, and therefore, no realistic cosmology can arise in this case (see e.g.~\cite{Bahamonde:2017ize}). 

\subsection{Dark sector couplings: Kinetic vs.~Potential effects} \label{sec:kin_pot}

In this section, we illustrate the cosmological evolution of the system for the different cases discussed earlier. By analysing various coupling scenarios, we highlight how kinetic and potential interactions influence the dynamics of dark energy and dark matter.

\subsubsection{Uncoupled Quintessence or the Illusion of No Coupling?}

\begin{figure}
        \centering
         \includegraphics[width=0.55\linewidth]{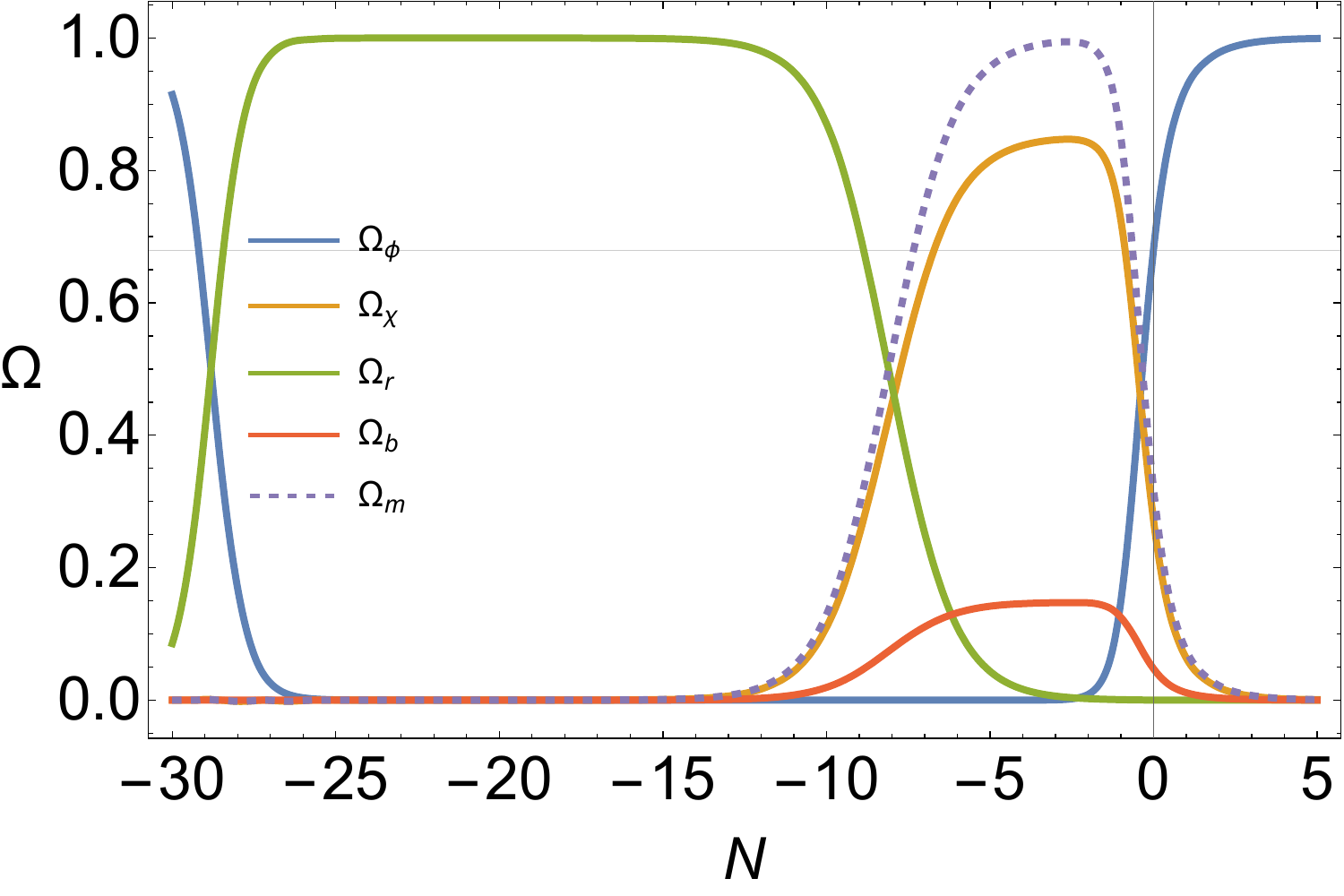}
        \caption{Energy densities' evolution for the system in \cref{eq:gen_dynsys_rad_bar}, with $\xi= 0$ and $\lambda=\sqrt{1.9}$. The initial conditions are given by $x(-15)= -10^{-6}$, $y(-15)= 1.013 \times 10^{-11}$, $v(-15)=0.9995$ and $u(-15)=0.01217$. These conditions produce a realistic cosmology ensuring a sufficiently long radiation domination epoch and $\Omega_{\phi,0}=0.681$, $\Omega_{\chi,0}=0.271$, $\Omega_{b,0}=0.0474$, $w_{\rm eff,0}=-0.483$, $w_{\phi,0}=-0.709$.}
        \label{fig:numsimxi=0}
    \end{figure}

We begin with \cref{fig:numsimxi=0}, which shows the evolution of the energy density rates for an example where \( 2\beta = -\gamma \) (i.e., \( \xi = 0 \)). In this case, the kinetic and potential couplings—though present— may cancel each other, creating the \textit{illusion} of no interaction between the scalar fields. This scenario effectively reduces to an uncoupled quintessence model at the background level, extensively studied in the literature (see e.g.~\cite{Bahamonde:2017ize}).  

\begin{figure}[H]
        \centering
        \includegraphics[width=0.50\linewidth]{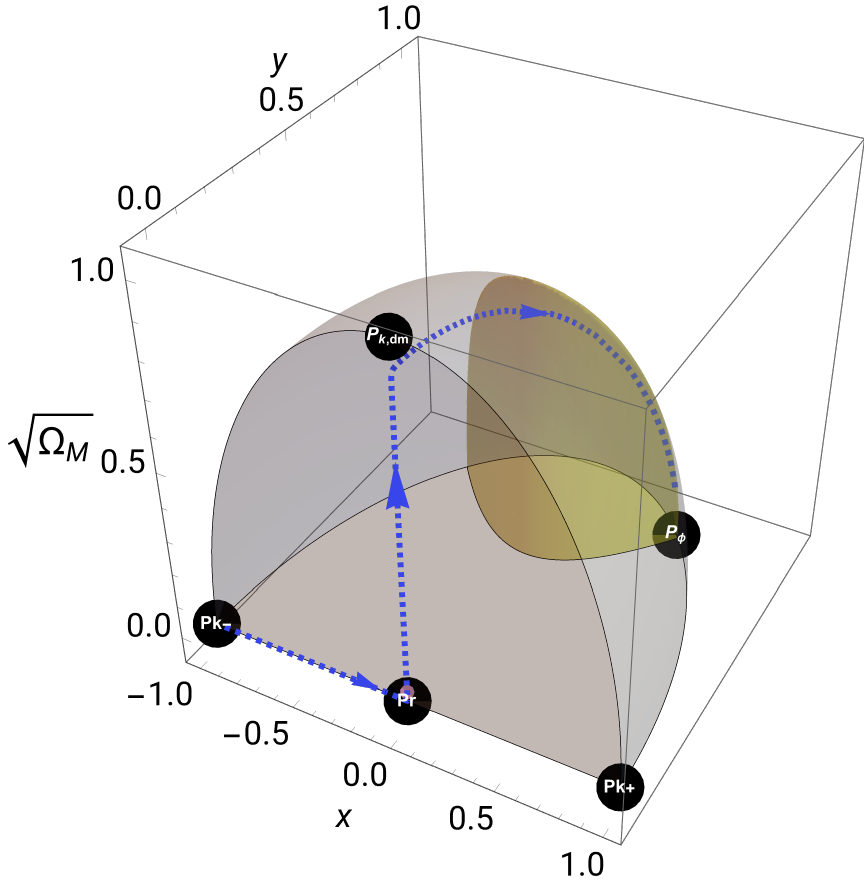}
        \hspace{0.05\linewidth}     
        \includegraphics[width=0.35\linewidth]{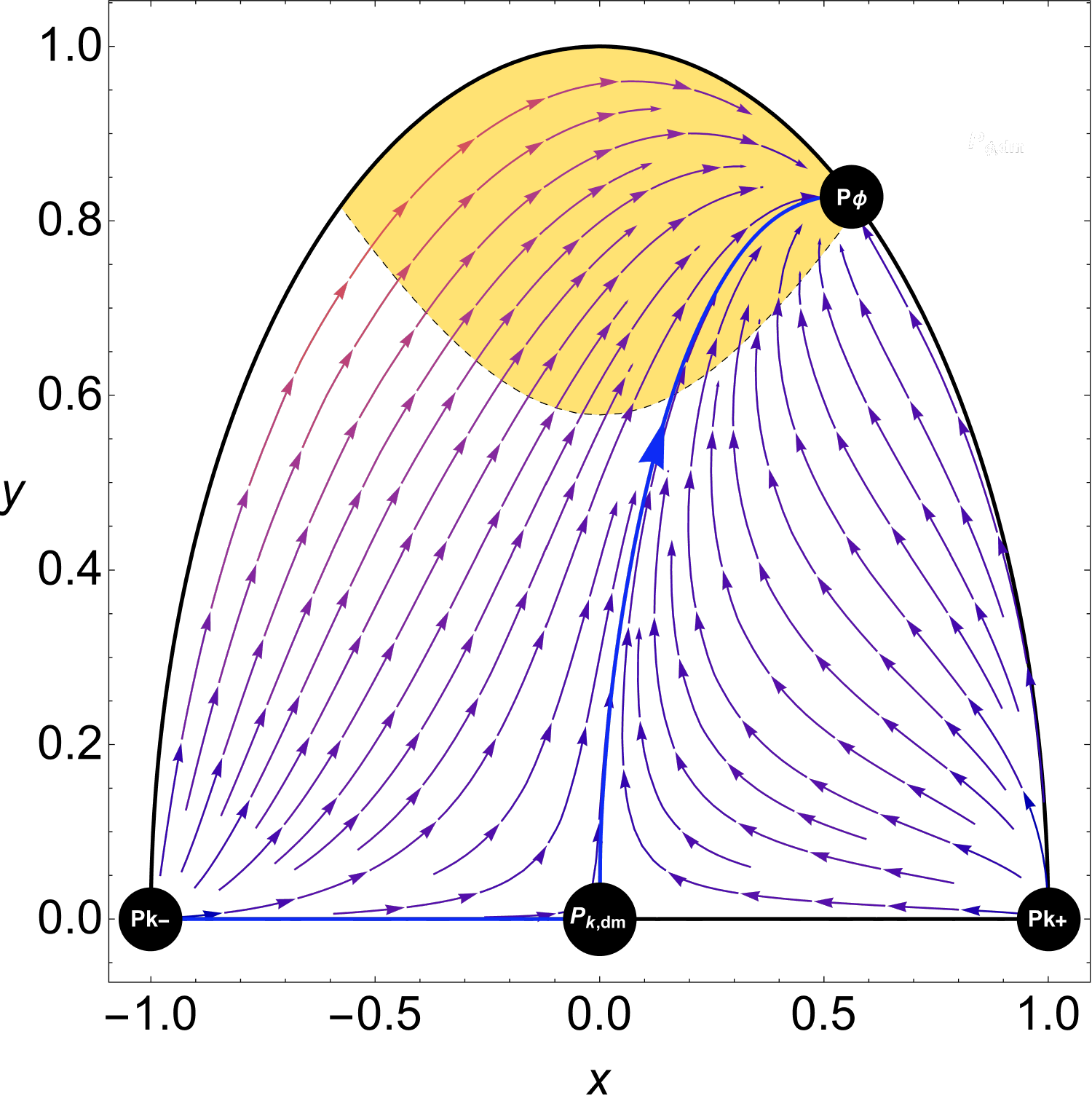}
            \caption{{\em Left panel:} The trajectory of system in \cref{fig:numsimxi=0} (dashed blue curve), embedded in a 3D phase space portrait. The dimensionless variable $\sqrt{\Omega_M}=\sqrt{1-x^2-y^2-v^2}$ represents the matter sector (dark matter and baryons). We show all existing fixed points except $P_{b}$ for clarity (very close to $P_{k,dm}$). The full phase space and the accelerating region are depicted in grey and yellow, respectively. {\em Right panel:} 2D phase portrait for $u=v=0$ representing the heteroclinic orbits for the same parameters as the simulation on the left. The projected orbit followed by the simulation in the 2D plane is depicted in blue.}   
         \label{fig:Falseabsentcoupling3D2D}
    \end{figure}

For illustrative purposes, we take \( \lambda < \sqrt{2} \), where the scalar-dominated fixed point \( P_\phi \) is an accelerating attractor. In the left panel of \cref{fig:Falseabsentcoupling3D2D}, 
we depict the trajectory of the system in \cref{fig:numsimxi=0}, embedded in a 3D phase space portrait showing variables $(x,y,\sqrt{\Omega_M})$, where $\Omega_M=\Omega_\chi+\Omega_b$ represents the full matter sector (dark matter and baryons). The full phase space and the accelerating region are depicted in grey and yellow, respectively. For this choice of parameters, the existing fixed points are $P_{k}^{\pm}$, $P_{r}$, $P_{b}$, $P_{k,dm}$, $P_\phi$. On the right panel, we show the 2D phase portrait for $u=v=0$ representing the heteroclinic orbits for the same parameter values as in \cref{fig:numsimxi=0}. In this plot, the blue line corresponds to the orbit followed by the simulation in the 2D plane. The initial conditions are chosen such that the present-day energy densities, \( \Omega_{i,0} \), are close to the observationally inferred values, as specified in \eqref{eq:EDs0}.  
Our choice of \( \lambda \) is larger than the best-fit value recently obtained for exponential quintessence, \( \lambda \lesssim0.537 \), \cite{Akrami:2018ylq, Bhattacharya:2024hep}, selected here solely to better illustrate the dynamics.

\subsubsection{Potential or Kinetic Coupling?}

\begin{figure}[H]
    \centering
    \includegraphics[width=0.55\linewidth]{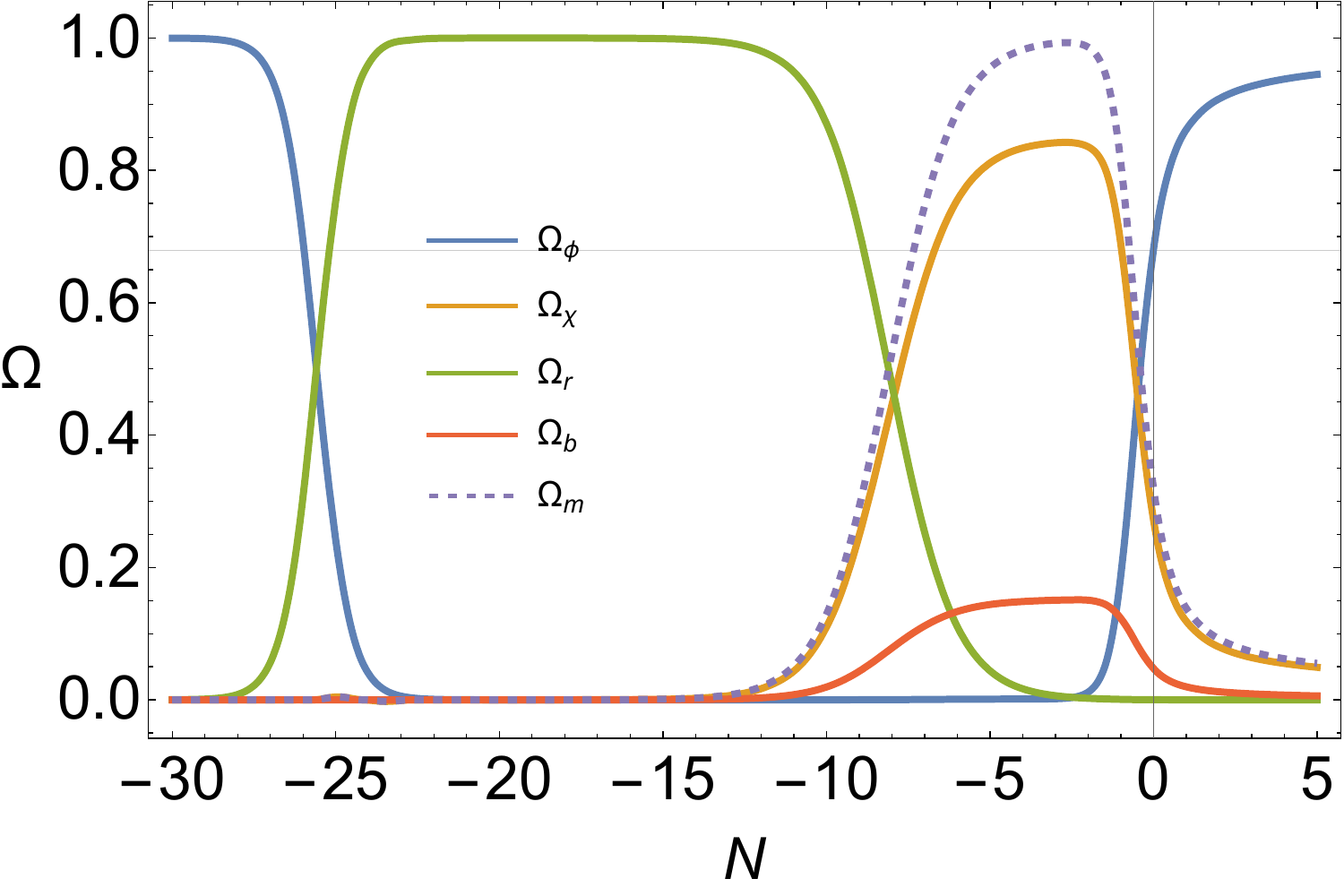}
    \caption{ Energy densities' evolution for the system in \cref{eq:gen_dynsys_rad_bar}, with $\xi= -0.1$ and $\lambda=\sqrt{3}$. The initial conditions are given by $x(-15)= -10^{-6}$, $y(-15)= 7.633 \times 10^{-12}$, $v(-15)=0.9995$ and $u(-15)=0.01221$. These conditions produce a realistic cosmology ensuring a sufficiently long radiation domination epoch and $\Omega_{\phi,0}=0.681$, $\Omega_{ \chi,0}=0.271$, $\Omega_{b,0}=0.0473$, $w_{\rm eff,0}=-0.373$, $w_{\phi,0}=-0.548$. }
    \label{fig:lambda=sqrt3DS+numSim}
\end{figure}

\begin{figure}[H]
    \centering
    \includegraphics[width=0.50\linewidth]{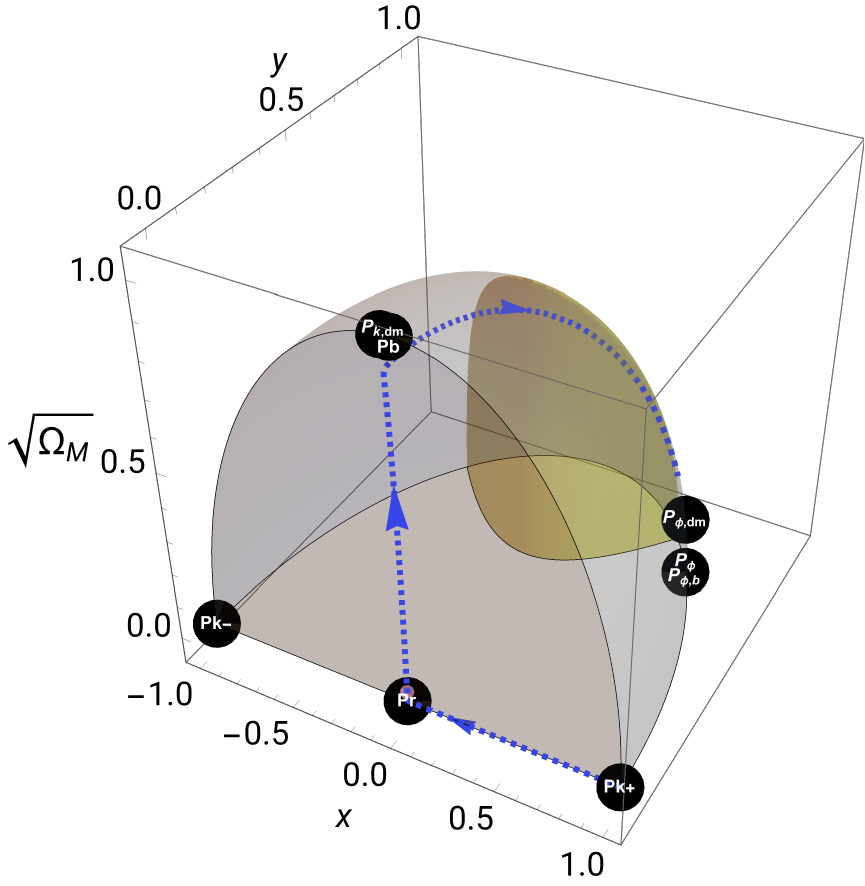}
    \hspace{0.05\linewidth}     
    \includegraphics[width=0.35\linewidth]{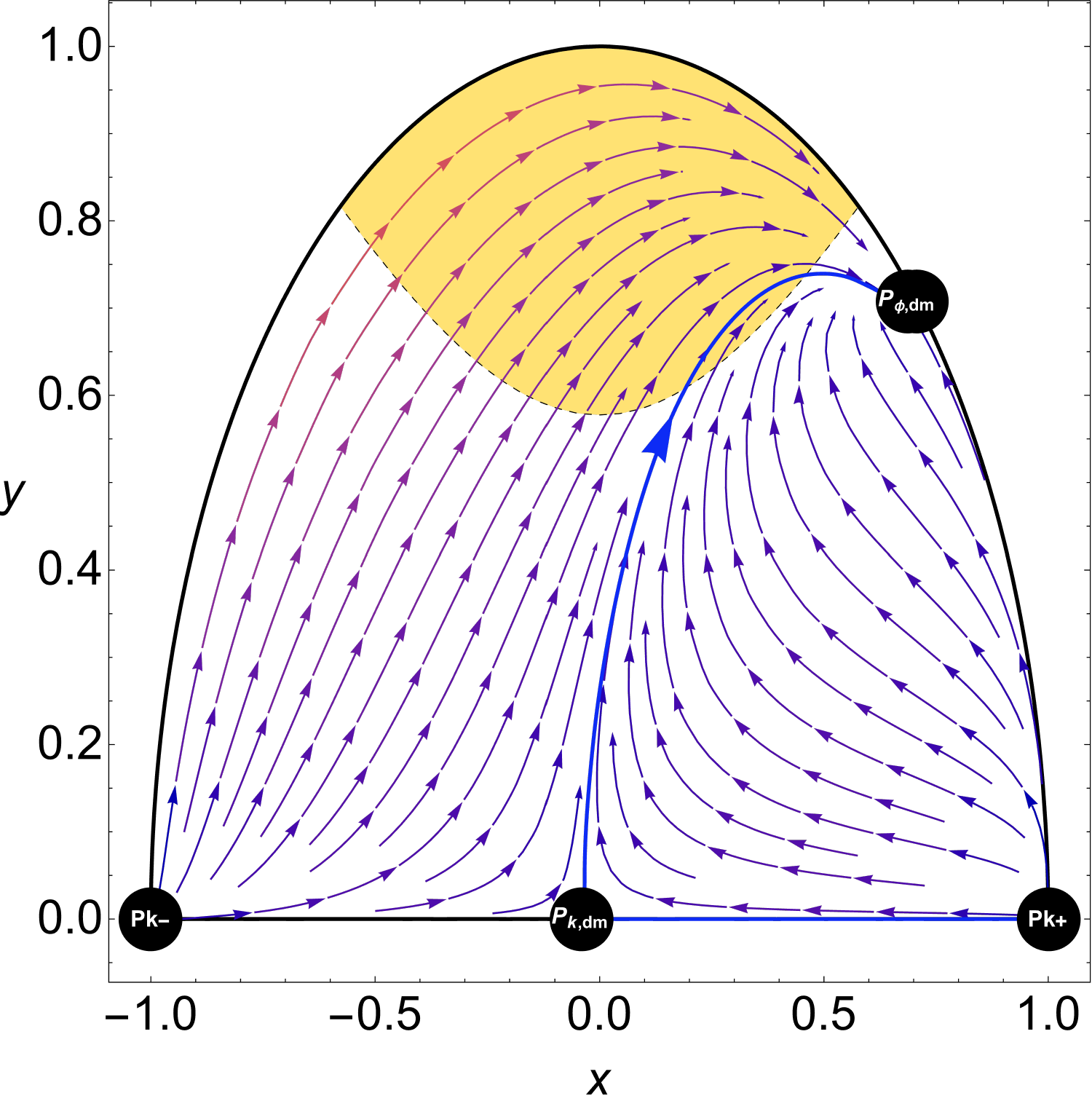}
        \caption{ {\em Left panel:} The trajectory of system in \cref{fig:lambda=sqrt3DS+numSim}, embedded in a 3D phase space portrait. The dimensionless variable $\sqrt{\Omega_M}=\sqrt{1-x^2-y^2-v^2}$ represents the matter sector (dark matter and baryons). We show all existing fixed points in black. The full phase space and the accelerating region are depicted in grey and yellow, respectively. {\em Right panel:} 2D phase portrait for $u=v=0$ representing the heteroclinic orbits for the same parameters as the simulation on the left. The projected orbit followed by the simulation in the 2D plane is depicted in blue.}
     \label{fig:lambda=sqrt3DS+numSim2}
\end{figure}

\begin{figure}[H]
    \centering
   \includegraphics[width=0.55\linewidth]{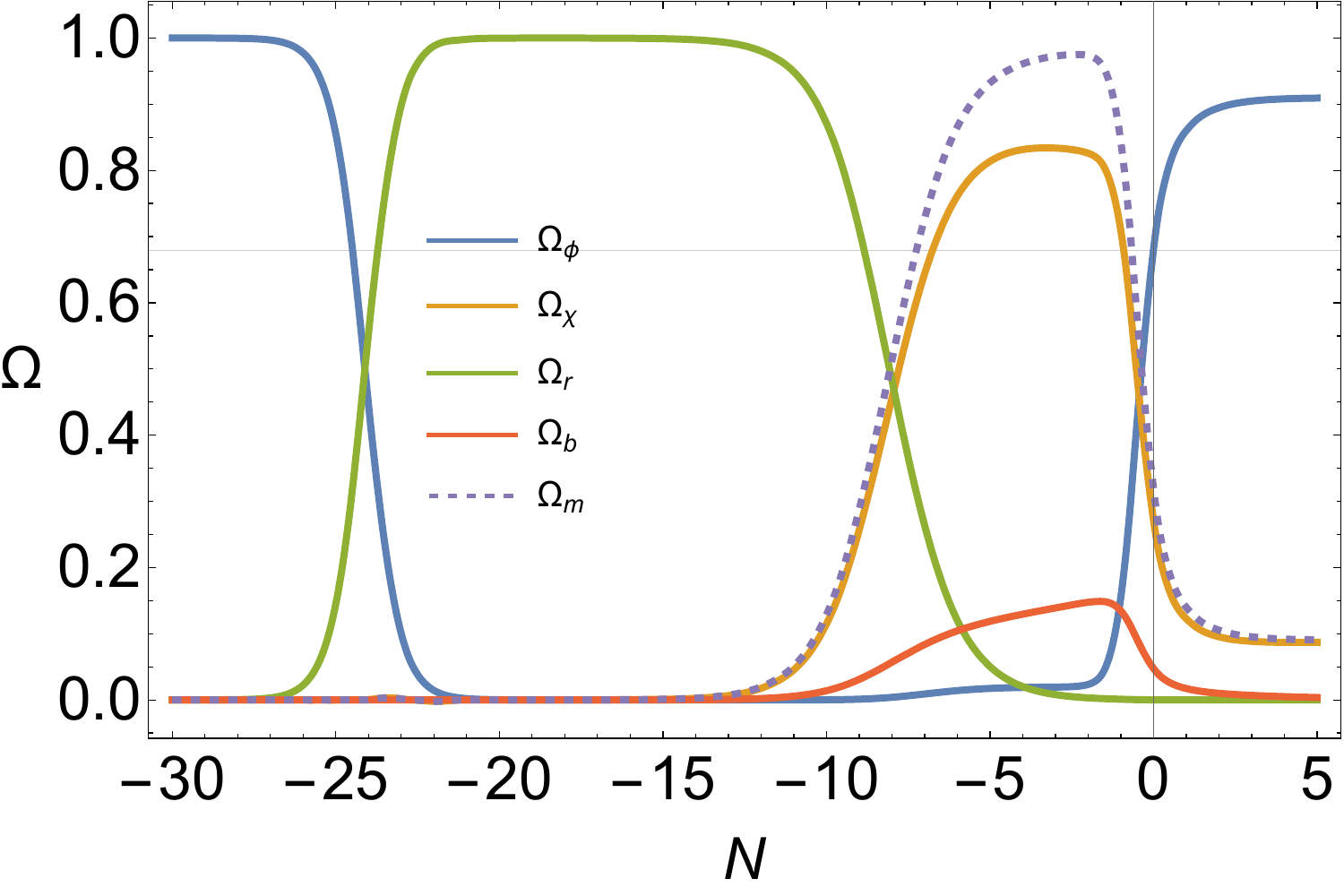} 
    \caption{Energy densities' evolution for the system in \cref{eq:gen_dynsys_rad_bar}, with $\xi= -0.4$ and $\lambda=\sqrt{3}$. The initial conditions are given by $x(-15)= 10^{-6}$, $y(-15)= 1.141 \times 10^{-12}$, $v(-15)=0.9995$ and $u(-15)=0.01035$. These conditions produce a realistic cosmology ensuring a sufficiently long radiation domination epoch and $\Omega_{\phi,0}=0.681$, $\Omega_{\chi,0}=0.271$, $\Omega_{b,0}=0.0474$, $w_{\rm eff,0}=-0.438$, $w_{\phi,0}=-0.642$.}
      \label{fig:lambda=sqrt3DS+numSim3}
\end{figure}

\begin{figure}[H]
    \centering
    \includegraphics[width=0.5\linewidth]{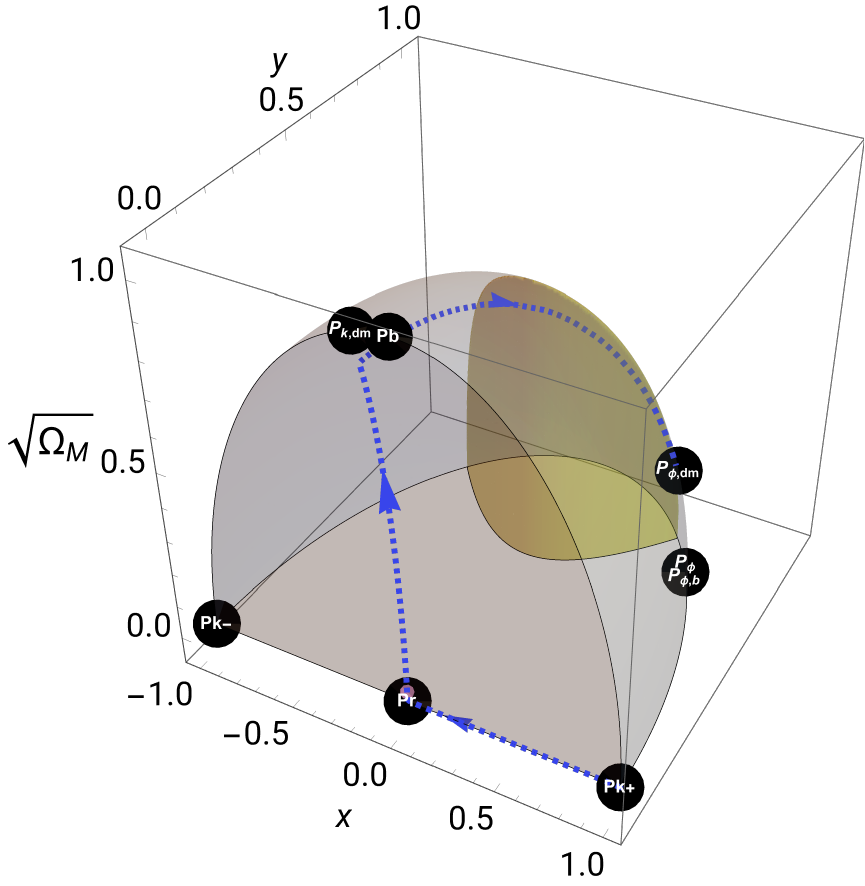}
    \hspace{0.05\linewidth}
    \includegraphics[width=0.35\linewidth]{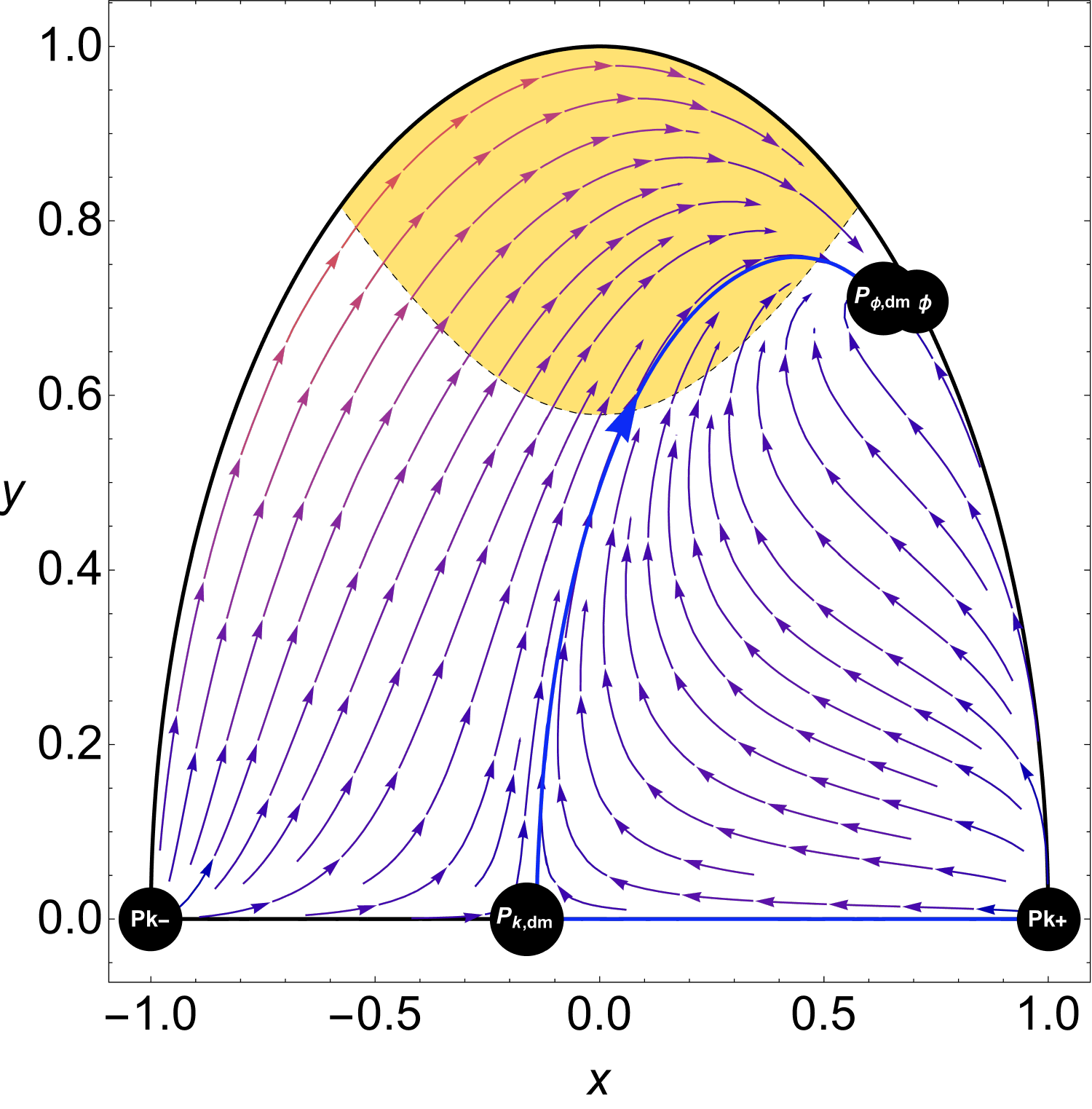}
        \caption{{\em Left panel:} The trajectory of system in \cref{fig:lambda=sqrt3DS+numSim}, embedded in a 3D phase space portrait. The dimensionless variable $\sqrt{\Omega_M}=\sqrt{1-x^2-y^2-v^2}$ represents the matter sector (dark matter and baryons). We show all existing fixed points in black. The full phase space and the accelerating region are depicted in grey and yellow, respectively. {\em Right panel:} 2D phase portrait for $u=v=0$ representing the heteroclinic orbits for the same parameters as the simulation on the left. The projected orbit followed by the simulation in the 2D plane is depicted in blue.}
    \label{fig:DSXi-0.4}
\end{figure}

In \cref{fig:lambda=sqrt3DS+numSim}-\ref{fig:DSXi-0.4}, we illustrate the cosmological evolution for \( \lambda = \sqrt{3} \), considering two different values of the kinetic--potential coupling, namely \( \xi = -0.1 \) and \( \xi = -0.4 \). For these parameter choices, the scaling fixed point \( P_{\phi,dm} \) emerges as the late-time attractor, while the scalar-dominated point \( P_\phi \) is a saddle and it coincides with \( P_{\phi,b} \).  Moreover, the fixed points $P_{\rm k}^{\pm}, P_r, P_b,$ also exist, as well as the (dark) matter scaling $P_{{\rm k},dm}$. 

In \cref{fig:lambda=sqrt3DS+numSim,fig:lambda=sqrt3DS+numSim3} we show the evolution of the energy density rates for these two examples. From these plots, we see that increasing the value of \( \xi \) significantly alters the evolution of the baryon energy density in a non-trivial manner, making the system observationally distinguishable from an uncoupled scenario. However, for sufficiently large \( |\xi| \), the scalar field starts to contribute non-negligibly during the matter-dominated era, imposing constraints on the maximal allowed value of \( \xi \).  

Our choice \( \lambda = \sqrt{3} \) is motivated by swampland conjectures, which suggest a lower bound \( \lambda > \sqrt{2} \) for exponential dark energy potentials \cite{Rudelius:2021oaz}. Consequently, the system exhibits only \emph{transient} cosmic acceleration rather than eternal acceleration, as reflected in the 3D phase space portrait of the trajectories shown in the left panels of \cref{fig:lambda=sqrt3DS+numSim2,fig:DSXi-0.4} as well as on the 2D phase space portraits on the right panels. Transient acceleration is particularly interesting as it naturally avoids the formation of eternal cosmological horizons, which pose challenges for defining a proper quantum gravity framework\footnote{Transient acceleration for exponential uncoupled quintessence for open universes has also been recently suggested as an alternative to eternal accelerating solutions \cite{Andriot:2024jsh}. However, observational constraints seem to rule out this possibility \cite{Bhattacharya:2024hep}}. 
In this scenario, the system evolves towards the scaling matter fixed point \( P_{\phi,dm} \), characterised by an effective equation of state  
\[
w_{\rm eff} = \frac{\xi}{2\lambda - \xi} \simeq (-0.03, -0.1)
\]
and a dark matter fractional density  
\[
\Omega_{\chi} \simeq (0.03, 0.09)
\]
for \( \xi = (-0.1, -0.4) \), respectively (see \cref{tab:gen_dynsys_rad_bar}), while $\Omega_\phi=1-\Omega_\chi$ at this fixed point.  While transient acceleration is disfavoured in the uncoupled case \cite{Akrami:2018ylq,Bhattacharya:2024hep}, the presence of the coupling may alter this conclusion in a non-trivial way. A full cosmological analysis is required to determine whether viable trajectories exist, which we leave for future work.

As discussed earlier, for the class of models considered—where both the field space metric and the scalar potential take exponential forms—the kinetic and potential couplings become \textit{degenerate} at the background level. This degeneracy implies that the cosmological evolution remains insensitive to whether the interaction is purely kinetic, purely potential, or a combination of both. However, perturbations and observational constraints may help break this degeneracy, potentially offering a way to distinguish between different coupling mechanisms.

\subsection{Axio-dilaton dark sector}\label{sec:AxioDil}

In this subsection, we analyse an axio-dilaton model recently studied in \cite{Smith:2024ayu,Smith:2024ibv}. This model explores a purely kinetic coupling between scalar dark matter and dark energy, corresponding to $\gamma=0$ in the interaction vector \cref{eq:Qsimple}, such that $\xi=2\beta$. Additionally, the model includes a coupling between dark energy and baryons, which we denote with $\mu$. However, this interaction is tightly constrained by observational bounds, including tests of the equivalence principle, fifth-force experiments, and constraints on the time variation of fundamental constants (see, e.g.~\cite{Will:2014kxa,Bertotti:2003rm}), and it is thus required to be very small.

Here, we revisit this system through the lens of dynamical systems, providing a structured and insightful analysis of its evolution. This approach not only clarifies the underlying dynamics but also highlights key similarities and distinctions compared to the coupled scalar systems studied earlier in this work. Moreover, it offers a deeper understanding of the results obtained in \cite{Smith:2024ayu,Smith:2024ibv}, situating them within a broader theoretical context.

\begin{figure}[H]
    \centering
    \includegraphics[width=0.4\linewidth]{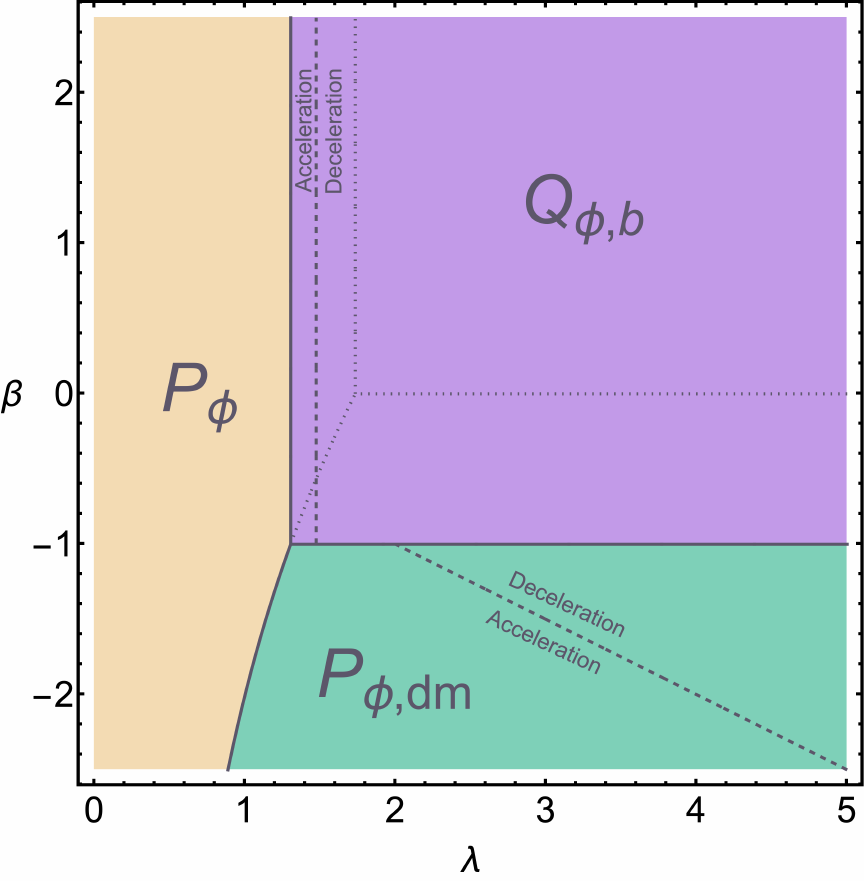} 
    \hspace{0.05\linewidth}
    \includegraphics[width=0.4\linewidth]{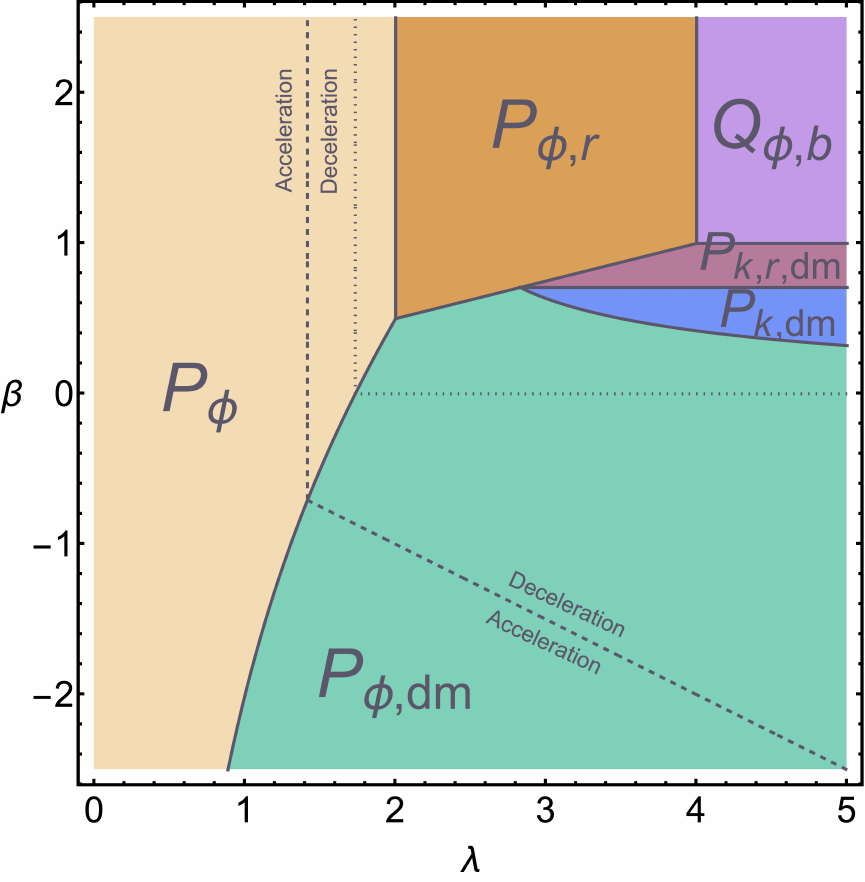}
    \caption{Illustration of the parameter space of the axio-dilaton system  \cref{eq:dynsys_axio_dil}, where each fixed point  as listed in \cref{tab:FPsAxioDilaton} is an attractor according to the study in \cref{tab:ADstabgen_r_b} for $\mu=1$ (left) and $\mu=-1$ (right). The dashed lines delimit the region where there is acceleration at the respective fixed point. The dotted lines show the boundaries for each fixed point in the case where $\mu=0$, as depicted in \cref{fig:reg_stab}. }\label{fig:reg_stab_axio_dil}
\end{figure}

In this model, dark matter and dark energy are both described by scalar fields, the {\em axio-dilaton} system. However, as we mentioned in \cref{sec:SDS}, in addition to its interaction with dark matter, the dark energy field (the dilaton) also couples to baryons  through a conformal coupling in the matter action:
$S_m(\psi, \tilde g)$,
where  $\tilde g_{\mu\nu} = C^2(\phi)g_{\mu\nu}$ with $C(\phi) = e^{\mu\phi}$. This gives rise to the following baryon-dilaton coupling
\cite{Smith:2024ibv, Smith:2024ayu}:
\be\label{eq:couplingB}
\dot\rho_b+3H\rho_b=\mu\rho_b\,\dot\phi \mathcomma
\ee
 where $\phi$ is the dilaton (dark energy). 
We use the same averaging approach as before and introduce the same variables as before in \cref{subsec:variables}. 
With those variables, the equations of motion become 
\begin{subequations}\label{eq:dynsys_axio_dil}
    \begin{align}
        x'=& -3\,x +3 \,x^3 +\frac{3}{\sqrt{6}} \lambda \,y^2+  2 \,x \,v^2   +\frac32 x\, u^2 \left(1-\sqrt{\frac{2}{3}}\mu \,x \right)+\frac32 x\,\Omega_\chi + \frac{\sqrt{6}}{2} \,\beta \, \Omega_\chi\mathcomma \\
        y'= & -\frac{\sqrt{6}}{2} \lambda x\,y + 3\, x^2\,y +2y\,v^2 +\frac32 y\, u^2 \left(1-\sqrt{\frac{2}{3}}\mu \,x \right) +\frac32 y\,\Omega_\chi \mathcomma\\
        v'= & -2 v + 3v\,x^2 + 2v^3 +\frac32 v\, u^2 \left(1-\sqrt{\frac{2}{3}}\mu \,x \right) +  \frac{3}{2} v\,\Omega_\chi \mathcomma\\
        u'= & -\frac32 u + 3 u\,x^2 +2 u\, v^2+\frac{\sqrt{6}}{2}\mu \,u\,x+ \frac{3}{2} u^3 \left(1-\sqrt{\frac{2}{3}}\mu \,x \right)  +  \frac{3}{2} u\,\Omega_\chi  \mathperiod
        \end{align}
\end{subequations}
Here, $\chi$ is the axion which plays the role of dark matter in the axio-dilaton system.
Thus, in this case, we have the couplings $(\mu, \beta, \lambda)$, associated with the coupling between dark energy and baryons, $\mu$, kinetic coupling between dark energy and dark matter, $\beta$ and the potential for dark energy $\lambda$, all of which are considered to be constant and thus the system \eqref{eq:dynsys_axio_dil} is closed. 
The new coupling, $\mu$, introduces two new fixed points, in addition to the fixed points arising in the case studied above, due to the coupling of dark energy with baryons. In  \cref{tab:FPsAxioDilaton}, we list these two new fixed points together with their properties. The stability of the complete set of fixed points for the axio-dilaton system \eqref{eq:dynsys_axio_dil} is summarised in \cref{tab:ADstabgen_r_b} and in \cref{fig:reg_stab_axio_dil}.   
The new fixed  points for the system are the following:
\begin{enumerate}[{\bf i.}]
    \item $Q_{{\rm k},b}$, {\em Kination-baryon scaling}. This point is new and arises due to the coupling of baryons with the dilaton. It does not give acceleration, and it is either a repeller or saddle (see \cref{tab:ADstabgen_r_b}). It exists for $\mu \geq 3/2$, and thus, it is not relevant for realistic cosmologies as such a large coupling will violate observational constraints. 
    \item $Q_{\phi,b}$, {\em Baryon scaling}. This point is dominated by dark energy $(x,y)\ne 0$ and baryons $u\ne0$. In the limit $\mu\to0$, this point coincides with $P_{\phi,b}$ in the uncoupled baryon case discussed before (see \cref{tab:gen_dynsys_rad_bar,tab:Stabgen_r_b}). It has $\Omega_b = 1-\frac{3}{\lambda(\lambda+\mu)}$  and an effective equation of state parameter $w_{\text{eff}} = -\frac{\mu}{\lambda+\mu}$ and therefore it can give rise to an accelerating universe for some values of the parameters. This point is, however, of little interest for realistic cosmologies as it exists for too large values of $\mu$. The stability properties of this point are summarised in  \cref{tab:ADstabgen_r_b} and \cref{fig:reg_stab_axio_dil}.

\end{enumerate}

\begin{table}[H]
    \centering
\resizebox{16.5cm}{!}{
\begin{tabular}{|c|c|c|c|c|c|}
\hline
\rowcolor{gray!30}   $(x,y,v,u)$ & $w_{eff}$ & Existence & $\Omega_\chi$ & Accelerating \\
    \hline 
   \ $Q_{{\rm k}, b} =\lp -\sqrt{\frac{3}{2}}\frac{1}{\mu}, 0, 0,\sqrt{1-\frac{3}{2\mu^2}} \rp$  & $\frac{3}{2\mu^2}$  &  $\mu^2\geq \frac32$ & $0$ & no \\
  \hline
  \ $Q_{\phi, b} =\lp \sqrt{\frac{3}{2}} \frac{1}{\lambda+\mu}, \sqrt{\frac{3}{2}}\sqrt{\frac{\lambda+2\mu}{\lambda(\lambda+\mu)^2}}, 0, \sqrt{1-\frac{3}{\lambda(\lambda+\mu)}} \rp$  & $-\frac{3\mu}{\lambda(\lambda+\mu)^2}$  & $0<\lambda \leq \sqrt{6}\land \mu \geq \frac{3-\lambda ^2}{\lambda }$ \  & $0$ & $0<\lambda <\sqrt{2} \ \land \ \frac{3-\lambda^2}{\lambda}\leq \mu < \frac{9-2\lambda^2}{2\lambda}+\frac{3}{2}\sqrt{-\frac{-9+4\lambda^2}{\lambda^2}}$\\ && $\lor$ \, & \, \, & $\lor$\, \\ && $\lambda >\sqrt{6}\ \land \ \mu \geq -\frac{\lambda}{2}$ && $\sqrt{2}\leq \lambda<\frac{3}{2} \ \land \ \frac{9-2\lambda^2}{2\lambda}-\frac{3}{2}\sqrt{-\frac{-9+4\lambda^2}{\lambda^2}}<\mu< \frac{9-2\lambda^2}{2\lambda}+\frac{3}{2}\sqrt{-\frac{9+4\lambda^2}{\lambda^2}}$ \ \\ 
  \hline
\end{tabular}
}
    \caption{New fixed points arising for system in \cref{eq:dynsys_axio_dil}. Note that for $\mu=0$ point  $Q_{{\rm k},b}$ does not exist while  $Q_{\phi,b}\to P_{\phi,b}$ in \cref{tab:gen_dynsys_rad_bar}. }
    \label{tab:FPsAxioDilaton}
    
\end{table}

\begin{table}[H]
    \centering
\resizebox{\textwidth}{!}{
\resizebox{16cm}{!}{
\begin{tabular}{|c|c|}
\hline
\rowcolor{gray!30}   Fixed Point & Stability \\
    \hline 
     $P_{\rm k}^{\pm}$ &  $P_{\rm k}^{-}$: Repeller for $\lambda \geq 0\land \beta >-\sqrt{\frac{3}{2}}\land \mu <\sqrt{\frac{3}{2}}$; Saddle otherwise\\
     &  $P_{\rm k}^{+}$: Repeller for $0\leq \lambda <\sqrt{6}\land \beta <\sqrt{\frac{3}{2}}\land \mu >-\sqrt{\frac{3}{2}}$; Saddle  otherwise \\
     \hline 
     $P_r$ & Saddle   \\
     \hline 
     $P_b$ & Saddle  \\
     \hline 
      $P_{{\rm k},dm}$ &   Attractor for $0<\beta <\frac{1}{\sqrt{2}}\land \mu <-\beta \land \lambda >\frac{2 \beta ^2+3}{2 \beta }$; Saddle otherwise  \\
     \hline
     $P_{{\rm k},r,dm}$  &  Attractor for $\left(\beta > \frac{1}{\sqrt{2}} \land \mu <-\beta \land \lambda >4 \beta \right)$; Saddle otherwise\\
     \hline
      $P_\phi$ & Attractor for $\lambda =0\lor \left(0<\lambda <2\land \beta >\frac{\lambda ^2-3}{\lambda }\land \mu <\frac{3-\lambda ^2}{\lambda }\right)$; Saddle otherwise \\
      \hline
  $P_{\phi, dm}$ & Attractor or saddle  for some values of the parameters. See \cref{fig:reg_stab_axio_dil} \\
  \hline
   $P_{\phi, r} $ &  Attractor for $\left(\lambda>2\land \beta >\frac{\lambda }{4}\land \mu <-\frac{\lambda }{4}\right)$; Saddle otherwise \\
      \hline
  $Q_{{\rm k}, b}$ & Repeller for $\mu<\frac{-\sqrt{6}}{2} \land \beta<-\mu \land \leq 0 \lambda<-2\mu  \lor \mu>\frac{\sqrt{6}}{2} \land \beta>-\mu \land \lambda \geq0  $; Saddle otherwise \\
  \hline
  $Q_{\phi, b}$ & Repeller, attractor or saddle for some values of the parameters. See \cref{fig:reg_stab_axio_dil} \\
   \hline
\end{tabular}
}}
    \caption{Stability of fixed points for system in \cref{eq:dynsys_axio_dil}. }
    \label{tab:ADstabgen_r_b}
\end{table}

\subsubsection{Background evolution}

\begin{figure}[H]
    \centering
    \includegraphics[width=0.55\linewidth]{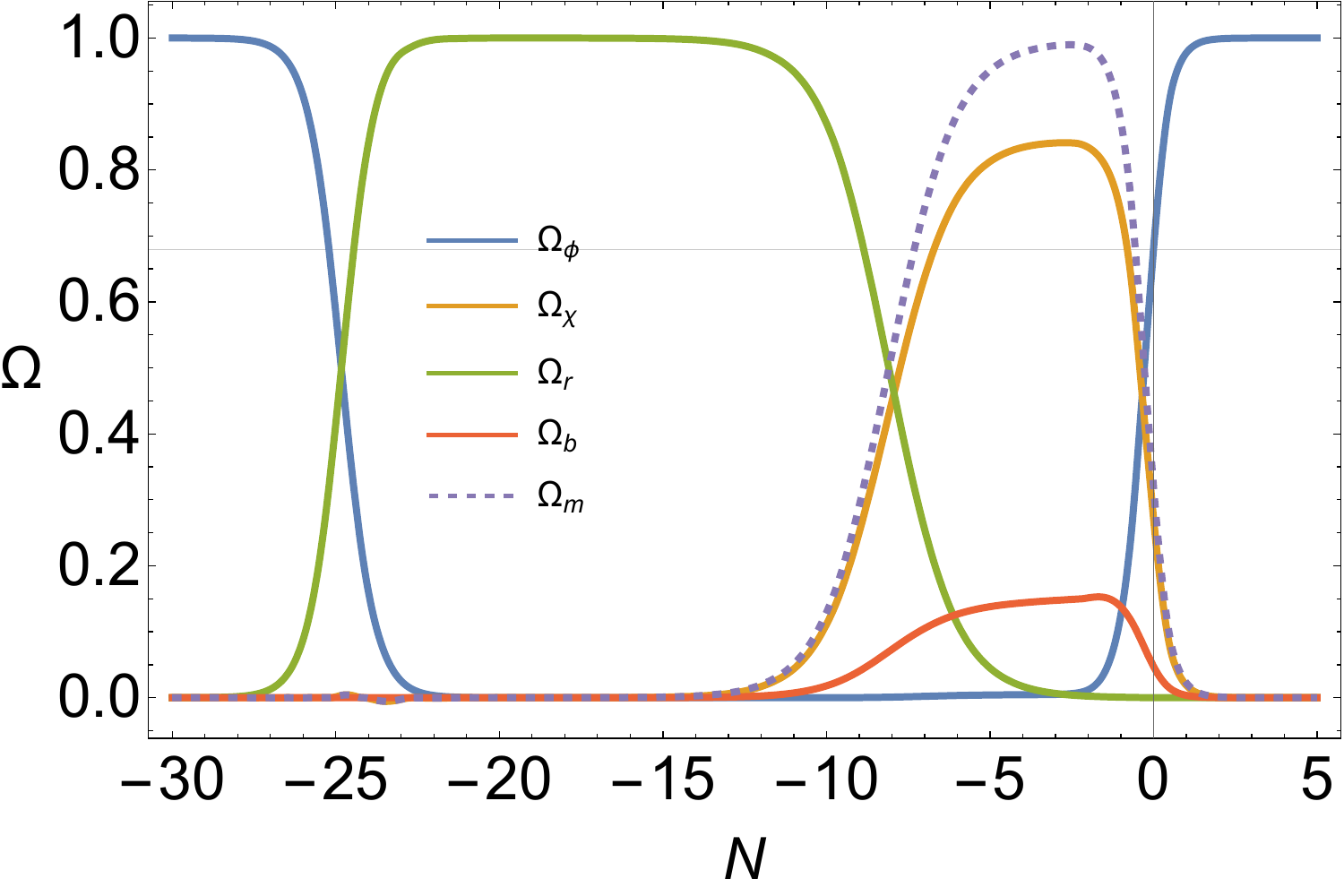}
    \caption{Energy densities' evolution for the axio-dilaton system, with $\mu=-10^{-3}$, $\beta= 0.1$ and $\lambda=0.4$. The boundary conditions are given by $x(-15)= -10^{-6}$, $y(-15)= 1.01709 \times 10^{-11}$, $v(-15)=0.9995$ and $u(-15)=0.0118$. These conditions produce a realistic cosmology ensuring a sufficiently long radiation domination epoch and $\Omega_{\phi,0}=0.681$, $\Omega_{\chi,0}=0.271$, $\Omega_{b,0}=0.0473$ $w_{\rm eff,0}=-0.651$, $w_{\phi,0}=-0.957$.}  
    \label{fig:axiodilaton}
\end{figure}

\begin{figure}[H]
    \centering
    \includegraphics[width=0.5\linewidth]{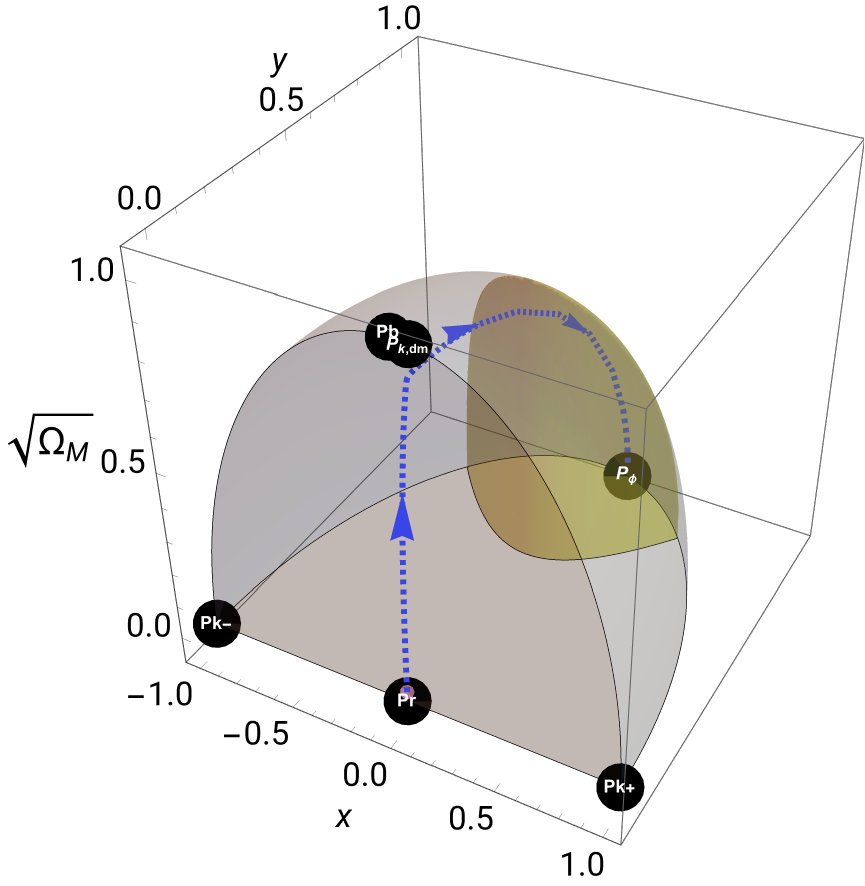}
    \hspace{0.05\linewidth}
    \includegraphics[width=0.35\linewidth]{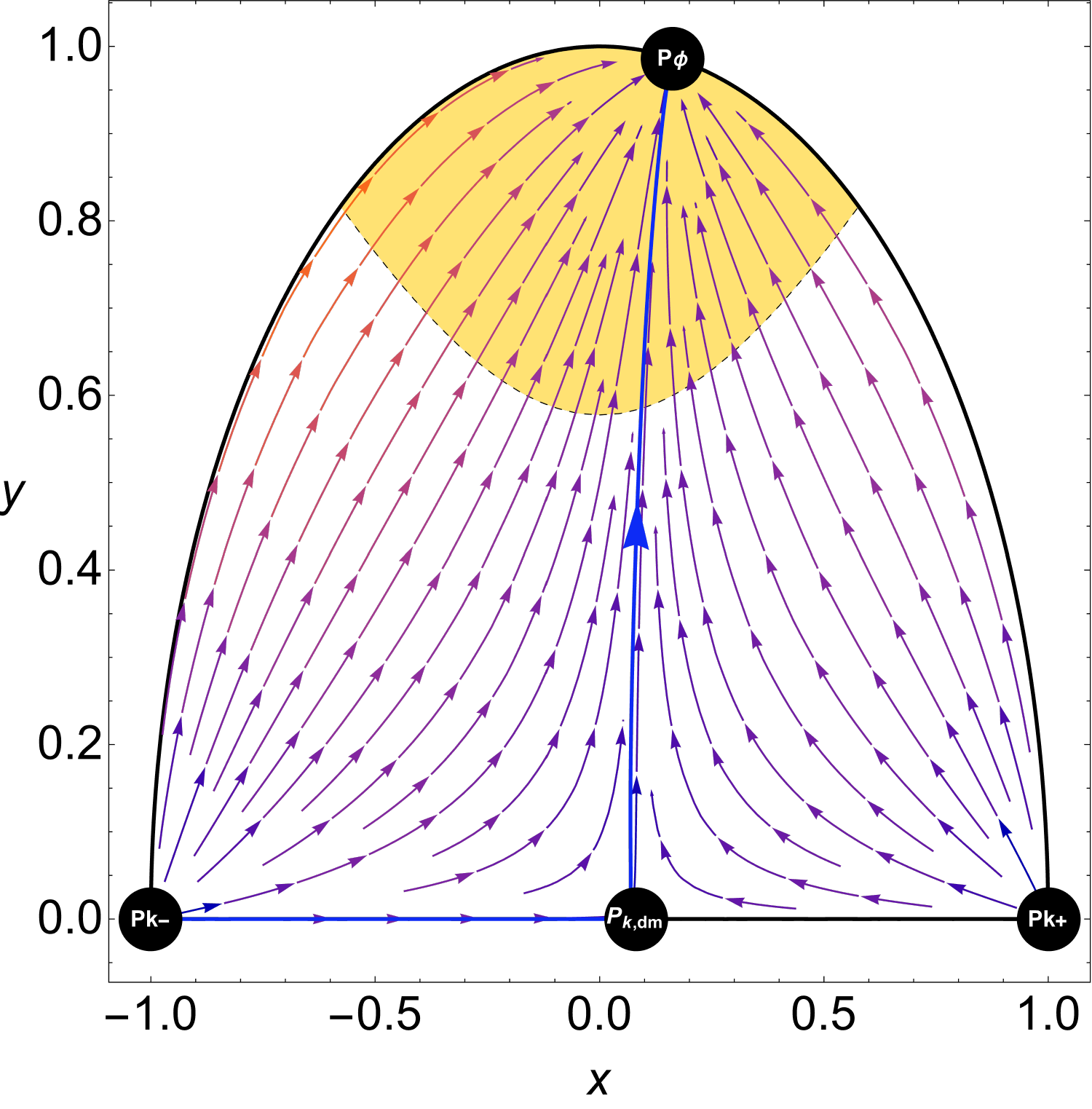}
    \caption{{\em Left panel:} The trajectory of system in \cref{fig:axiodilaton} (dashed blue curve), embedded in a 3D phase space portrait. The dimensionless variable $\sqrt{\Omega_M}=\sqrt{1-x^2-y^2-v^2}$ represents the matter sector (dark matter and baryons). We show all existing fixed points except $P_{b}$ for clarity (very close to $P_{k,dm}$). The full phase space and the accelerating region are depicted in grey and yellow, respectively. {\em Right panel:} 2D phase portrait for $u=v=0$ representing the heteroclinic orbits for the same parameters as the simulation on the left. The projected orbit followed by the simulation in the 2D plane is depicted in blue.}
    \label{fig:axiodilaton3d}
\end{figure}

We now look at the cosmological evolution of the axio-dilaton system considered in \cite{Smith:2024ayu,Smith:2024ibv}. The authors considered the following phenomenological values for the parameters:
\be\label{eq:paramAD}
\lambda = 4\beta \lesssim 0.5 \mathcomma \quad \mu =-10^{-3} \mathcomma
\ee
where the first constraint is motivated by having dark energy domination today driven by the dilaton with $w_{\phi}\lesssim -0.9$, while the value for $\mu$ is motivated by observational constraints mentioned above on light scalars. From the dynamical system analysis, we see that for these values, none of the new fixed points in \cref{tab:FPsAxioDilaton} exists, while the attractor is the standard quintessence point, $P_\phi$, driving the universe towards eternal acceleration. On the other hand, dark matter is driven by the $P_{{\rm k},dm}$ kination-matter scaling point, with $\beta\in (0,0.1)$ (this corresponds to $\xi=2\beta=(0,0.2)$). In \cref{fig:axiodilaton} we illustrate the evolution of the energy density rates for the parameter values within the range \eqref{eq:paramAD},  $(\beta,\lambda,\mu)=(0.1,0.4,-10^{-3})$. Further in the left panel of \cref{fig:axiodilaton3d}, we show the corresponding trajectory in the 3D phase space (dashed blue curve) and in the right panel, the projected 2D phase portrait for $u=v=0$ with the heteroclinic orbits for the same parameters showing in blue the orbit followed by the simulation in the 2D plane. As we can see, the evolution starts in the far past at kination, moving through radiation, matter domination and eternal acceleration in the future. 

\section{Discussion}\label{sec:conclu}

Coupled dark sector models are a natural extension of the $\Lambda$CDM model, allowing for a dynamical form of dark energy that drives the current accelerated expansion while also relaxing the assumption of no interaction between dark energy and dark matter.

In string theory-inspired cosmological models, axions and their companions — the saxions (or moduli) — provide a natural realisation of coupled scalar dark sectors. In these setups, interactions between the fields generically arise both from their kinetic terms, via a non-trivial field space metric, and from their potential energy. Crucially, both types of coupling are typically present and cannot be arbitrarily switched off.

While axions and saxions have been considered separately as candidates for dark energy (see e.g.~\cite{Bhattacharya:2024kxp}), in this work, we explored the more natural scenario where both fields evolve dynamically, with one acting as dark matter and the other as dark energy. More generally, purely field-theoretic coupled scalar sectors can feature both kinetic and potential interactions.

We showed how these couplings can be elegantly encoded in an interaction vector, $Q^\nu$ (see eq.~\eqref{eq:KPIntQ}), commonly used in coupled dark sector models, clarifying its physical origin. This formulation makes the individual roles of kinetic and potential interactions explicit, allowing us to analyse their effects systematically.

By identifying the axion with dark matter and the saxion with dark energy--following the oscillation averaging procedure of \cite{Sa:2021eft} -- we applied the robust dynamical systems approach to study the system's cosmological evolution. For concreteness, we considered exponential forms for the field space metric, the potential coupling, and the saxion's potential. A key result of our analysis is that, in this case, kinetic and potential couplings become degenerate: their distinction disappears at the level of the background evolution. As a result, a cancellation between these couplings can render the system effectively uncoupled, even though interactions can be present at the fundamental level.

Focusing on realistic cosmological trajectories that reproduce the observed abundances of radiation, dark matter, baryons, and dark energy, we found that the kinetic-potential coupling parameter, $\xi$ (see eq.~\eqref{eq:xidef}), can be relatively large. Furthermore, for an exponential saxion potential, the accelerated expansion can either be eternal -- if $\lambda< \sqrt{2}$, leading to the standard quintessence attractor—or transient-- if $\lambda>\sqrt{2}$, where a (dark) matter scaling fixed point $P_{\phi,dm}$ emerges as the new attractor. Whether this latter trajectory is compatible with the most recent observational constraints requires a full cosmological analysis, which we leave for future work.

Beyond background evolution, an important open question is whether the degeneracy between kinetic and potential couplings can be observationally broken. While large-scale structure surveys such as DESI \cite{DESI:2024mwx} and supernova surveys such as DESY5 \cite{DES:2024tys}, which primarily probe the background evolution, may be insufficient to distinguish them, additional constraints from perturbations or non-linear effects could help disentangle and constrain these parameters separately. In particular, it would be interesting to derive observational bounds on the field space metric $f(\phi)$, which is typically exponential in supergravity and string theory models (see e.g.~the recent work on saxion-axion multifield inflation in supergravity \cite{Aragam:2021scu,Bhattacharya:2022fze}). Although we focused on the simplest exponential case, our interacting vector $Q^{\nu}$ framework allows for the study of more general forms of the metric and potentials.

Finally, we revisited the recently proposed axio-dilaton model of \cite{Smith:2024ayu,Smith:2024ibv} within our dynamical systems framework. This model effectively reduces to a purely kinetic-coupled scenario, a special case of our general analysis. While the model includes an additional coupling between the dilaton and baryons, this interaction is highly constrained by observational bounds, eliminating the two additional fixed points that would otherwise arise. Furthermore, our results demonstrate that this purely kinetic-coupled model is indistinguishable from scenarios with both kinetic and potential couplings, provided that both interactions are governed by exponential functions of the scalar field.

Our findings open several avenues for further investigation. A full cosmological analysis, including the evolution of perturbations, is essential to assess these coupled scalar models' observational viability and determine whether the kinetic potential degeneracy can be broken. Additionally, it would be interesting to explore more general forms of the field space metric and potentials beyond the simple exponential case considered here. Finally, a deeper connection with explicit UV completions, such as string theory compactifications, may provide further theoretical constraints on the coupling parameters and their cosmological implications.

\section*{Acknowledgments}
We thank David J.E.~Marsh (Doddy), Maria Mylova, Vivian Poulin, and Gianmassimo Tasinato for useful discussions. The work of SR and IZ is partially supported by STFC grants ST/T000813/1 and ST/X000648/1. 
EMT is supported by funding from the European Research Council (ERC) under the European Union’s HORIZON-ERC-2022 (grant agreement no. 101076865). \\
For the purpose of open access, the authors have applied a Creative Commons Attribution license to any Author Accepted Manuscript version arising. 
Data access statement: no new data were generated for this work.

\addcontentsline{toc}{section}{References}
\bibliographystyle{utphys}

\bibliography{refsCDS}

\end{document}